\documentclass[12pt]{article}
\usepackage[a4paper, top= 3cm,bottom=3cm, left=0.8 in, right=0.8 in]{geometry}

\usepackage{amsmath}
\usepackage{amsthm}
\usepackage{amssymb}
\usepackage{graphicx}
\usepackage{color}
\usepackage{enumerate}
\usepackage{natbib}
\usepackage{url}
\usepackage{multirow}
\usepackage{lineno}
\usepackage{makecell}
\usepackage{ulem}
\usepackage{xr}
\usepackage{xr-hyper}
\usepackage{subcaption}
\usepackage{booktabs}
\usepackage[ruled,vlined,linesnumbered]{algorithm2e}
\usepackage{float}
\usepackage{setspace}
\usepackage{titlesec}
\titleformat{\section}
  {\Large\bfseries}
  {\thesection}                
  {1em}             
  {}

\titlespacing*{\paragraph}
  {0pt}    
  {0pt}   
  {1em}

\usepackage{etoolbox}
\AtBeginEnvironment{thebibliography}{
    \small  
    \singlespacing
}

\AtBeginEnvironment{align}{
  \setlength{\abovedisplayskip}{5pt}     
  \setlength{\belowdisplayskip}{5pt}     
  \setlength{\abovedisplayshortskip}{3pt} 
  \setlength{\belowdisplayshortskip}{3pt} 
}
\AtBeginEnvironment{align*}{
  \setlength{\abovedisplayskip}{5pt}     
  \setlength{\belowdisplayskip}{5pt}     
  \setlength{\abovedisplayshortskip}{3pt} 
  \setlength{\belowdisplayshortskip}{3pt}
}

\AtBeginEnvironment{equation}{
  \setlength{\abovedisplayskip}{5pt}     
  \setlength{\belowdisplayskip}{5pt}     
}
\AtBeginEnvironment{equation*}{
  \setlength{\abovedisplayskip}{5pt}     
  \setlength{\belowdisplayskip}{5pt}    
}

\usepackage{hyperref}
\hypersetup{
    colorlinks=true,
    linkcolor=black,
    filecolor=magenta,      
    urlcolor=cyan,
    citecolor = blue,
    pdftitle={Overleaf Example},
    pdfpagemode=FullScreen,
    }

\usepackage{cleveref}

\newtheorem{theorem}{\textbf{Theorem}}

\newtheorem{definition}{Definition}

\newtheorem{proposition}{Proposition}
\newtheorem{assumption}{Assumption}

\newcommand{\bm}{\boldsymbol}

\def\0{\boldsymbol 0}

\newcommand{\blind}{1}

\makeatletter
\newcommand*{\addFileDependency}[1]{
  \typeout{(#1)}
  \@addtofilelist{#1}
  \IfFileExists{#1}{}{\typeout{No file #1.}}
}
\makeatother

\renewenvironment{abstract}
 {
  \begin{center}
  \bfseries \abstractname\vspace{-.5em}\vspace{-.5em}
  \end{center}
  \list{}{
    \setlength{\leftmargin}{0pt}
    \setlength{\rightmargin}{\leftmargin}
    \small
  }
  \item\relax}
 {\endlist}

\makeatletter
\def\@maketitle{
  \begin{center}
  \let \footnote \thanks
    {\large \@title \par}
    {
      \begin{tabular}[t]{c}
        \@author
      \end{tabular}\par}
  \end{center}
  \par
  }
\makeatother

\begin{document}

\if1\blind
{
  \title{\bf A Unified Principal Component Analysis for Stationary Functional Time Series}
  \author{Zerui Guo\hspace{.2cm}\\
    School of Mathematics, Sun Yat-sen University, Guangzhou, Guangdong, China\\
    Jianbin Tan \thanks{
    Jianbin Tan is the co-first author.}\\
    Department of Biostatistics and Bioinformatics, Duke University, Durham, NC 27708, USA\\
    and \\
    Hui Huang\thanks{Email of correspondence: \texttt{huangh89@ruc.edu.cn}} \\
    Center for Applied Statistics and School of Statistics, Renmin University of China, Beijing, China}
    \date{}
  \maketitle
} \fi

\if0\blind
{
  \bigskip
  \bigskip
  \bigskip
  \begin{center}
    {\LARGE\bf A Unified Principal Components Analysis for Stationary Functional Time Series}
\end{center}
  \medskip
} \fi

\begin{sloppypar}
\begin{abstract}

Functional time series (FTS) data have become increasingly available in real-world applications. Research on such data typically focuses on two objectives: curve reconstruction and forecasting, both of which require efficient dimension reduction. While functional principal component analysis (FPCA) serves as a standard tool, existing methods often fail to achieve simultaneous parsimony and optimality in dimension reduction, thereby restricting their practical implementation. To address this limitation, we propose a novel notion termed optimal functional filters, which unifies and enhances conventional FPCA methodologies. Specifically, we establish connections among diverse FPCA approaches through a dependence-adaptive representer for stationary FTS. Building on this theoretical foundation, we develop an estimation procedure for optimal functional filters that enables both dimension reduction and prediction within a Bayesian modeling framework. Theoretical properties are established for the proposed methodology, and comprehensive simulation studies validate its superiority over competing approaches. We further illustrate our method through an application to reconstructing and forecasting daily air pollutant concentration trajectories.

\end{abstract}

\noindent
{\it Keywords:} Functional time series, 
dynamic functional principal component analysis, 
optimal functional filter, 
dependency adaptivity.

\clearpage

\section{Introduction}

Serially dependent functional data have become pervasive in modern applications such as traffic flow analysis \citep{ma2024network}, stock price forecasting \citep{chang2023autocovariance}, and air pollution monitoring \citep{wang2023nonlinear, tan2024graphical}. These data are commonly referred to as functional time series  \citep[FTS;][]{bosq2000linear, hormann2010weakly, hormann2015estimation, rubin2020functional}, for which analysis tasks often involve curve reconstruction and forecasting, necessitating the capture of dependencies both within and between random functions.
Functional principal component analysis (FPCA) is a widely employed technique to accomplish these tasks \citep{hyndman2009forecasting, hormann2015dynamic, aue2015prediction, klepsch2017prediction, tan2024graphical, chang2024modelling}, offering various frameworks for handling dependencies within FTS. However, existing FPCA methods typically rely on rigid assumptions and consequently lack dependence-adaptive formulations. This often leads to either oversimplified representations or redundant models, limiting their practical utility.

Conventional FPCA methods, rooted in Karhunen-Lo\`eve (KL) expansions \citep{hsing2015theoretical, wang2016functional}, focus on within-curve dependencies defined by lag-0 covariance functions. This approach enables dimension reduction by representing curves as a linear combination of static eigenfunctions and uncorrelated scores \citep{yao2005functional,hsing2015theoretical}. 
Eigenfunctions and FPC scores can be estimated using pre-smoothing \citep{ramsay2005functional} or pooling-smoothing techniques \citep{yao2005functional, li2010uniform, zhou2022theory}. Subsequent forecasting often relies on time series models for the scores, such as autoregression (AR) \citep{hyndman2009forecasting, aue2015prediction}, autoregressive moving average (ARMA) \citep{klepsch2017prediction}, or factor models \citep{ma2024network}. 
However, since these methods derive eigenfunctions and scores solely from lag-0 covariances, they inherently neglect serial dependencies across functional observations. This limitation renders conventional FPCA overly simplistic and theoretically suboptimal for dimension reduction in FTS \citep{panaretos2013cramer, hormann2015dynamic}.

Recent advancements in FTS analysis have sought to address serial dependencies through dynamic functional principal component analysis (DFPCA), which incorporates lag-$h$ covariances ($h \geq 1$) to capture temporal interdependencies across functional observations \citep{bathia2010identifying, hormann2015dynamic, gao2019high, tang2022clustering, chang2024modelling}. Notably, \citet{hormann2015dynamic}  introduced a generalized DFPCA framework that uses spectral density operators in the frequency domain to fully capture auto-correlation patterns. Their methodology constructs a dynamic KL expansion through the convolution of functional filters and dynamic FPC scores, where functional filters are obtained from eigen-decomposition of the spectral density operator. This formulation achieves theoretically optimal representations for stationary FTS \citep{hormann2015dynamic, tan2024graphical}.

However, the DFPCA proposed by \citet{hormann2015dynamic} faces critical limitations that hinder its direct applicability to FTS forecasting: First, its reliance on dense functional observations restricts its applicability to sparsely and irregularly sampled data \citep{kuenzer2021principal}. Second, the estimation of dynamic FPC scores requires future curve information, introducing operational infeasibility and potential bias in forecasting contexts \citep{aue2015prediction, koner2023second}. Third, and most critically, the non-uniqueness of spectral eigenfunctions induces substantial variability in functional filters and scores \citep{hormann2015dynamic, tan2024graphical}, often resulting in redundant representations that undermine predictive efficiency. These constraints highlight the need for a more adaptable and parsimonious framework to reconcile theoretical optimality with practical implementation.

In this work, we propose a unified framework of principal component analysis for stationary FTS, designed to provide a dependency-adaptive dimension reduction. The term ``unified'' has multiple implications in this context. First, we bridge conventional FPCAs and dynamic FPCAs under a cohesive theoretical foundation, enabling systematic exploration of serial dependence structures inherent to diverse dimension reduction paradigms. Second, it integrates the complementary strengths of these approaches, namely, optimality and parsimony. Third, it introduces a data-adaptive mechanism that autonomously selects the optimal representation form (static or dynamic) based on the intrinsic dependency structure of FTS, ensuring theoretical optimality without compromising predictive efficiency.

\begin{figure}[ht]
\captionsetup{width=\linewidth}
\centering
		\includegraphics[width=0.75\linewidth]{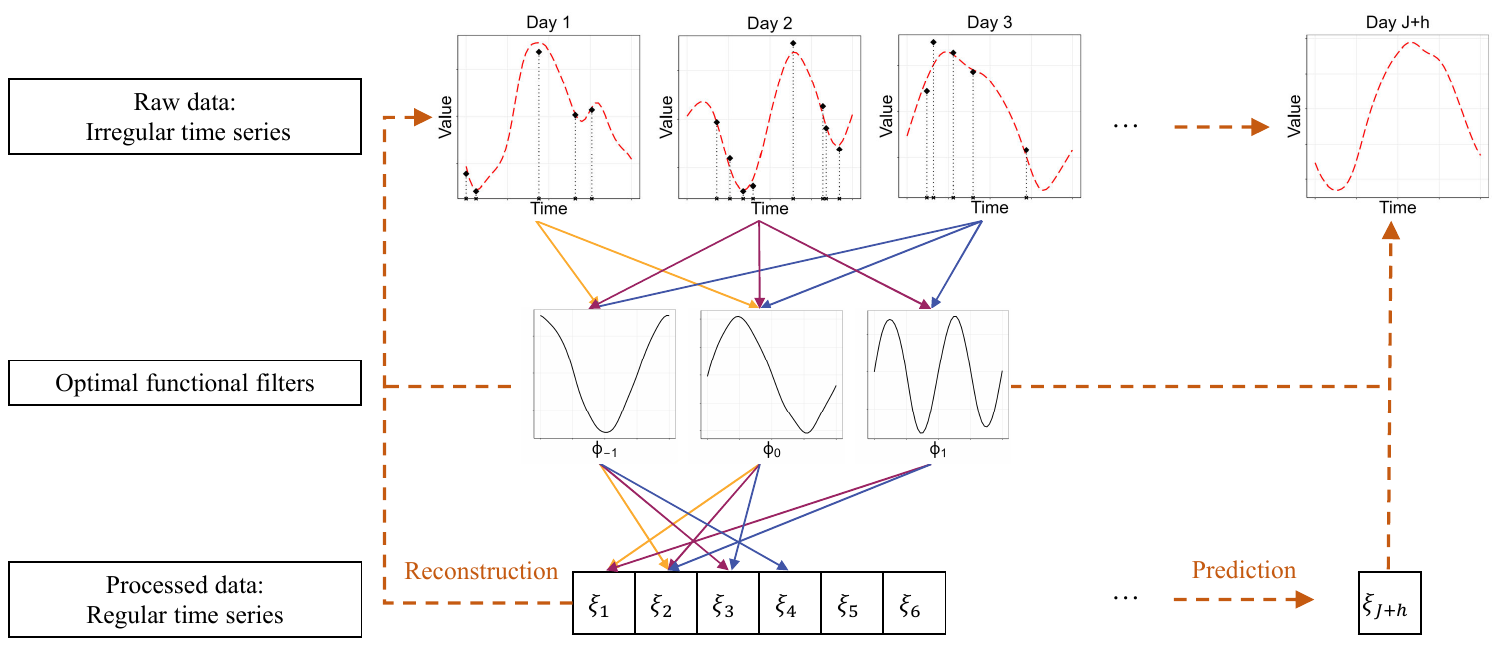}
	\caption{A flow chart of PADA.}
\label{flowchart}
\end{figure}

To achieve these objectives, we introduce \textbf{optimal functional filters}, as illustrated in Figure~\ref{flowchart}, a novel concept grounded in the DFPCA framework of \citet{hormann2015dynamic}.
These filters enable the construction of parsimonious dynamic KL expansions by adaptively balancing within-curve and between-curve dependencies. The adaptivity relies on weak separability, a condition characterizing covariance structures in dependent functional data \citep{liang2022test, zapata2022partial, tan2024graphical}. Under weak separability, the optimal functional filters collapse to conventional KL eigenfunctions, recovering static FPCA as a special case. In contrast, when dependencies violate weak separability, the filters yield sparse dynamic KL expansions tailored for forecasting. More importantly, this transition is fully automated, requiring no prior assumption or validation of weak separability. Our framework ensures that dimension reduction optimally aligns with the dependency structure of the FTS, advancing both theoretical coherence and practical utility.
Therefore, we refer to this framework as \textbf{Principal Analysis via Dependency-Adaptivity (PADA)}

The remainder of this article is organized as follows: In Section \ref{section2}, we establish a theoretical unification of existing FPCA and DFPCA methodologies through the lens of optimal functional filters. Section \ref{section3} details the proposed PADA procedure, including the estimation of optimal functional filters, dynamic FPC scores, and how to use them for curve reconstruction and forecasting. In Section \ref{ST}, we establish the statistical consistency of the optimal functional filters. Section \ref{section4} presents comprehensive simulation studies to validate the efficacy of PADA under diverse dependency structures. In Section \ref{section_case_study}, we demonstrate the practical use of our method through an application to PM2.5 concentration data, reconstructing and forecasting pollutant trajectories. Finally, Section \ref{section_discussion} concludes with a discussion of broader implications, limitations, and future research directions.

\section{A Unified Framework for FPCA and DFPCA} \label{section2}

Denote $L^2\left(\mathcal{T}, \mathbb{C}\right)$ as the Hilbert space of square-integrable complex-valued functions on $\mathcal{T}$, equipped with the inner product $\langle f, g\rangle:=\int_{\mathcal{T}} \overline{f(t)} g(t) \mathrm{d} t$ and the $L^2$ norm $\|f\|:=\langle f, f\rangle^{1 / 2}$, where $f, g \in L^2\left(\mathcal{T}, \mathbb{C}\right)$, $\overline{(\cdot)}$ is the conjugate of a complex number. For convenience, we assume that $\mathcal{T} = [0,1]$ in what follows. Moreover, $|\boldsymbol{\alpha}|$ is the length of a complex vector $\boldsymbol{\alpha}$ defined as $\sqrt{\boldsymbol{\alpha}^{*} \boldsymbol{\alpha}}$, where $(\cdot)^{*}$ is the conjugate transpose operation on a complex-valued vector.

Consider a FTS $\left\{X_j(\cdot); j \in \mathbb{Z}\right\}$, where $j \in \mathbb{Z}$ is a discrete-time index and each $X_j(\cdot)$ is a random element in $L^2\left([0,1], \mathbb{R}\right)$. We assume that $\left\{X_j(\cdot); j \in \mathbb{Z}\right\}$ is weakly stationary, i.e., 
\begin{center}
    $\operatorname{E}\{X_j(\cdot)\}$ and $\operatorname{Cov}\left\{X_{j+h}(\cdot), X_j(\cdot)\right\}$
\end{center}
is free of $j$ for any $h\in \mathbb{Z}$. This condition is commonly assumed for FTS analysis in literature \citep{bathia2010identifying, hormann2015dynamic, paparoditis2023bootstrap}. We define $\mu(t):= \operatorname{E}\{X_j(t)\}$ and $c_h(t, s):= \operatorname{Cov}\left\{X_{j+h}(t), X_j(s)\right\}$ for $t,s \in [0,1]$ and $j,h \in \mathbb{Z}$, which are called the mean and auto-covariance functions of $\left\{X_j(\cdot); j \in \mathbb{Z}\right\}$, respectively.

Under stationarity, we define the spectral density functions $\{f(\cdot, \cdot \mid \omega); \omega \in [-\pi, \pi]\}$ of FTS. Specifically, we apply the Fourier transform to $\{c_h(\cdot, \cdot); h \in \mathbb{Z}\}$
\begin{align}
f(t, s \mid \omega)=\frac{1}{2 \pi} \sum_{h \in \mathbb{Z}} c_h(t, s) \exp (\mathrm{i} h \omega), \quad t, s \in[0,1],\ \omega \in[-\pi, \pi],
\label{f to r}
\end{align}
where $\mathrm{i}$ denotes the imaginary unit. To ensure convergence, we impose a summability condition on $\{c_h(\cdot, \cdot); h \in \mathbb{Z}\}$:
$\sum_{h \in \mathbb{Z}}\left\{\int_0^1 \int_0^1\left|c_h(t, s)\right|^2 \mathrm{~d} t \mathrm{~d} s\right\}^{1 / 2}<\infty$.
As such, $\{c_h(\cdot, \cdot); h \in \mathbb{Z}\}$ can be represented using the inverse Fourier transform of $f(\cdot, \cdot \mid \omega)$:
\begin{align}
c_h(t, s)=\int_{-\pi}^\pi f(t, s \mid \omega) \exp (-\mathrm{i} h \omega)\ \mathrm{d} \omega, \quad t, s \in[0,1],\ h \in\mathbb{Z}. \label{r to f}
\end{align}
For each $\omega \in [-\pi, \pi]$, it can be shown that $f(\cdot, \cdot \mid \omega)$ is a positive-definite kernel, and thus by Mercer’s theorem \citep{hsing2015theoretical}, $f(\cdot, \cdot \mid \omega)$ admits a decomposition as
\begin{align}
f(t, s \mid \omega)=\sum_{k=1}^{\infty} \eta_k(\omega) \overline{\psi_k(t \mid \omega)} {\psi_k(s \mid \omega)},  \quad t,s \in [0,1], \ \omega \in [-\pi, \pi], \label{spectral decomp}
\end{align}
where $\eta_k(\omega)$ and $\psi_k(\cdot\mid \omega)\in L^2([0,1],\mathbb{C})$ are the $k$th eigenvalue and eigenfunction of $f(\cdot,\cdot \mid \omega)$, respectively. Here, $\psi_k(\cdot \mid \omega)$ is unique up to multiplication by a function $\nu_k(\cdot)$ on the complex unit circle $\mathcal{M}:=\big\{\nu:[-\pi,\pi] \rightarrow \mathbb{C}; |\nu(\omega)| = 1, \nu(\omega) = \overline{\nu(-\omega)}\big\}$. In other words, $\psi_k(\cdot \mid \omega)\nu_k(\omega)$ is always the $k$th eigenfunction of $f(\cdot,\cdot \mid \omega)$ given any $\nu_k(\cdot)\in \mathcal{M}$ and $\omega\in [-\pi,\pi]$.

We define a zero-mean FTS as $\varepsilon_j(t) :=X_j(t)- \mu(t)$, for $t \in [0,1]$ and $j \in \mathbb{Z}$, and simplify $\{\varepsilon_j(\cdot);j \in \mathbb{Z}\}$ as $\{\varepsilon_j;j \in \mathbb{Z}\}$ for convenience.  
The KL expansion for $\varepsilon_j$ is then given by
\begin{align}
    \varepsilon_j(t)=\sum_{k=1}^{\infty} \varphi_{k}(t)\xi_{jk} \quad  \text{with} \quad \xi_{jk} = \langle \varepsilon_{j}, \varphi_{k}\rangle, \quad t \in [0,1], \label{SFPCA fts}
\end{align}
where $\varphi_k(\cdot)$ is the $k$th eigenfunction of the lag-0 covariance function $c_0(\cdot,\cdot)$ and $\{\xi_{jk};j \in \mathbb{Z}\}$ are the FPC scores.
This expansion is a fundamental basis for conventional FPCA method \citep{yao2005functional, hsing2015theoretical, wang2016functional}, where the FPC scores satisfy $\operatorname{cov}(\xi_{jk_1}, \xi_{jk_2}) = 0$ whenever $k_1 \neq k_2$ for each $j \in \mathbb{Z}$. Nonetheless, it is possible for $\xi_{j_1k_1}$ and $\xi_{j_2k_2}$ to exhibit temporal correlations when $j_1\neq j_2$, regardless of the choice of $k_1$ and $k_2$ \citep{aue2015prediction}. As a result, the serial dependencies of $\left\{\varepsilon_j; j \in \mathbb{Z}\right\}$ can be inherited by the FPC scores.

\subsection{Optimal Functional Filters}\label{Op_FF}
It is worth noting that the KL expansion~\eqref{SFPCA fts} models $\varepsilon_j$ solely through its own FPC scores $\big\{\xi_{jk};k\geq 1\big\}$ for each $j \in \mathbb{Z}$. This kind of static expansion might be overly simplistic and not optimal for capturing serial dependencies among $\{\varepsilon_j; j \in \mathbb{Z} \}$.
Alternatively, \citet{hormann2015dynamic} proposed the dynamic KL expansion for $\varepsilon_j$:
\begin{align}
\varepsilon_j(t)=\sum_{k=1}^{\infty}\sum_{l \in \mathbb{Z}} \phi_{kl}(t)\xi_{(j+l)k} \quad
        \text{with} \quad \xi_{jk} = \sum_{l \in \mathbb{Z}} \langle \varepsilon_{j-l}, \phi_{kl}\rangle, \quad t \in [0,1], 
        \label{dyn kl}
\end{align}
where $\big\{\phi_{kl}(\cdot);l \in \mathbb{Z}\big\}$ is called the $k$th functional filters, given by
\begin{align}
    \phi_{kl}(t) = \frac{1}{2\pi} \int_{-\pi}^{\pi}\psi_k(t \mid \omega) \exp (-\operatorname{i}l\omega) \mathrm{d} \omega, \quad t \in [0,1], \ l \in \mathbb{Z}, \label{function filter}
\end{align}
and $\{\xi_{jk}; j \in \mathbb{Z}\}$ is a stationary time series with the spectral density $\eta_k(\cdot)$ in \eqref{spectral decomp}, refer to as the dynamic FPC scores.
Compared to \eqref{SFPCA fts}, the $k$th eigenfunction $\varphi_k(\cdot)$ is substituted by the $k$th functional filters $\big\{\phi_{kl}(\cdot);l \in \mathbb{Z}\big\}$ in \eqref{dyn kl}. 
Consequently, the dynamic FPC scores like $\xi_{(j-1)k}$ and $\xi_{(j+1)k}$ can contribute to the modeling of $\varepsilon_j$ for each $j$, allowing a more flexible representation of FTS while capturing serial dependencies.
Besides, the series $\{\xi_{jk}; j \in \mathbb{Z}\}$ in \eqref{dyn kl} are always uncorrelated across different components $k$ \citep{hormann2015dynamic}, and \eqref{dyn kl} is an optimal dimension reduction for $\{\varepsilon_j; j \in \mathbb{Z}\}$ in the sense of minimizing $L^2$ norm, i.e., $\forall K>0$,
\begin{align}
    \operatorname{E}\big\|\varepsilon_j - \sum_{k=1}^K\sum_{l \in \mathbb{Z}}\phi_{kl}(\cdot)\xi_{(j+l)k}\big\|^2 \leq \operatorname{E}\big\|\varepsilon_j - \sum_{k=1}^K\sum_{l \in \mathbb{Z}}\tilde{v}_{kl}(\cdot)\tilde{\xi}_{(j+l)k}\big\|^2,
    \label{opt reconstruction}
\end{align}
with $\tilde{\xi}_{jk} = \sum_{l \in \mathbb{Z}} \langle \varepsilon_{j-l}, \tilde{w}_{kl}\rangle$, where $\big\{\tilde{v}_{kl}(\cdot); l \in \mathbb{Z}\big\}$ and $\big\{\tilde{w}_{kl}(\cdot); l \in \mathbb{Z}\big\}$ are any two sequences of functions in $L^2\left([0,1], \mathbb{R}\right)$; see Theorem 2 in \citet{hormann2015dynamic} for more details.
{By taking $\tilde{v}_{k0}(t) = \tilde{w}_{k0}(t) = \varphi_k(t)$ and $\tilde{v}_{kl}(t) = \tilde{w}_{kl}(t) = 0$ when $l\neq 0$, $\forall t\in [0,1]$ and $k\geq 1$, \eqref{opt reconstruction} indicates that the dynamic KL expansion \eqref{dyn kl} is better than the static KL expansion~\eqref{SFPCA fts}} in the sense of minimizing $L^2$ norm.

While the dynamic KL expansion is theoretically optimal, it may diverge the temporal signal inherited by its FPC scores in an arbitrary way. To see this, we recall that the $k$th eigenfunctions $\psi_k(\cdot \mid \omega)$ of $f(\cdot,\cdot\mid \omega)$ in \eqref{spectral decomp} can be altered by different multiplicative factors $\nu_k(\cdot)\in\mathcal{M}$. This property leads to varying outputs of the functional filters $\big\{\phi_{kl}(\cdot);l \in \mathbb{Z}\big\}$ given by \eqref{function filter}. 
Although \citet{hormann2015dynamic} demonstrated that the reconstruction in \eqref{dyn kl} is unique given any valid functional filters, there may exist functional filters $\big\{\phi_{kl}(\cdot);l \in \mathbb{Z}\big\}$ for which $\sup_{l\in \mathbb{Z}}\|\phi_{kl}\|$ is arbitrarily small. 
If this happens, each dynamic FPC score in $\{\xi_{jk};j\in \mathbb{Z}\}$ contributes to $\varepsilon_j$  in a negligible way. In other words, we need lots of scores to recover the major pattern of $\varepsilon_j$. This is unrealistic in practice when we adopt a truncated version of \eqref{dyn kl} for modeling FTS:
\begin{align}
    \sum_{k=1}^K\sum_{|l|\leq L_k}\phi_{kl}(\cdot)\xi_{(j+l)k},
    \label{finite_dyn_kl}
\end{align}
where $L_k$s are some finite truncations. This approach, as applied in \citet{hormann2015dynamic}, may require large $L_k$s in \eqref{finite_dyn_kl} and thereby lead to a bloated representation for FTS.

To address the above issue, we first investigate the theoretical properties of the space of all valid functional filters.
Let $\mathcal{A}_k$ be the collection of all valid $k$th functional filters, i.e., if $\big\{\phi_{kl}(\cdot);l\in\mathbb{Z}\big\} \in \mathcal{A}_k$, its Fourier transformation 
$
\sum_{l\in \mathbb{Z}}\phi_{kl}(\cdot)\exp(\operatorname{i}l\omega)
$
is the $k$th eigenfunction of $f(\cdot,\cdot\mid \omega)$, $\forall$ $\omega\in [-\pi,\pi]$.
Denote $\big\{\phi_{kl}(\cdot);l\in\mathbb{Z}\big\}$ as $\Phi_k$, define $\|\Phi_k\|_{2}:=\sqrt{\sum_{l\in \mathbb{Z}}\|\phi_{kl}\|^2}$ and $\|\Phi_{k}\|_{\infty}:=\sup_{l\in \mathbb{Z}}\|\phi_{kl}\|$ as the $L^2$ and $L^{\infty}$ norms for $\Phi_{k}\in \mathcal{A}_k$, respectively. 

\begin{proposition}[Properties of the Space of Valid Functional Filters]
For stationary FTS $\{\varepsilon_j; j \in \mathbb{Z}\}$, $\mathcal{A}_k$ has the following properties:
\begin{itemize}
    \item[\textnormal{(a)}] For any $\Phi_k \in \mathcal{A}_k$, $\mathcal{T}_{h}(\Phi_{k})\in \mathcal{A}_k$ with $||\mathcal{T}_{h}(\Phi_{k})||_{2}=||\Phi_{k}||_{2}$ and $||\mathcal{T}_{h}(\Phi_{k})||_{\infty}=||\Phi_{k}||_{\infty}$, where $\mathcal{T}_{h}$ is an operation that shifts $\big\{\phi_{kl}(\cdot);l \in \mathbb{Z}\big\}$ to $\big\{\phi_{k(l+h)}(\cdot);l \in \mathbb{Z}\big\}$, $h \in \mathbb{Z}$. 
    \item[\textnormal{(b)}] For any $\Phi_{k}\in \mathcal{A}_k$, $||\Phi_{k}||_{2}=1$ and $||\Phi_{k}||_{\infty}\in [0,1]$.
    \item[\textnormal{(c)}] For any $\Phi_k\in \mathcal{A}_k$, there exists a $\phi_{kl}(\cdot)\in \Phi_k$ s.t. $\|\Phi_k\|_{\infty}=\|\phi_{kl}\|$.
\end{itemize}
\label{proposition2}
\end{proposition}

Proof of Proposition \ref{proposition2} can be found in the Supplementary Materials.

According to Proposition~\ref{proposition2}~(a), functional filters obtained by the shifting operation not only share the same \( L^2 \) and \( L^{\infty} \) norms but are also contained within \( \mathcal{A}_k \). These functional filters can be considered as the same element in $\mathcal{A}_k$; we refer to this phenomenon as the shift-invariant property of $\mathcal{A}_k$.

Note that the term $\exp(-\operatorname{i} l \omega)$ in \eqref{function filter} is a smooth function of $l$, the norm $\|\phi_{kl}\|$ typically varies smoothly as $l$ changes. Consequently, a large norm $\|\Phi_k\|_{\infty}$ suggests that there exists a sequence of functions in $\Phi_k$ that have relatively large norms.
Since $\|\Phi_k\|_2^2 = \sum_{l \in \mathbb{Z}} \|\phi_{kl}\|^2 = 1$ (see Proposition \ref{proposition2} (b)), the norms of functions that are not included in the sequence will shrink to $0$ as $\|\Phi_k\|_{\infty}$ increases. In other words, $\big\{\|\phi_{kl}\|; l \in \mathbb{Z}\big\}$ concentrates on a sequence of functions in $\Phi_k$ as its $L^{\infty}$ norm increases.
By this observation, we utilize $\|\Phi_k\|_{\infty}$ to measure the concentration degree of the functional filters $\Phi_k$.

Based on the above thoughts, we consider an optimization problem
\begin{align}
    \tilde{\Phi}_{k} = \mathop{\arg\max}\limits_{\Phi_{k} \in \mathcal{A}_k}||\Phi_{k}||_{\infty}, \quad k \geq 1\label{opt functional filter},
\end{align}
where $\tilde{\Phi}_{k} := \big\{\tilde{\phi}_{kl}(\cdot);l\in\mathbb{Z}\big\}$ is called the $k$th \textbf{optimal functional filters} for FTS. 
By design, optimal functional filters achieve the most parsimonious representation of FTS through the dynamic KL expansion. By maximizing the $L^{\infty}$ norms in \eqref{opt functional filter}, our method ensures norms of functional filters are concentrated on the fewest possible lagged components. This concentration is critical in practice, where FTS modeling relies on finite truncations as shown in equation \eqref{finite_dyn_kl}. For a fixed number of components  $K$, the truncated scores $\{\xi_{(j-L_k)k}, \ldots, \xi_{(j+L_k)k}\}$ inherit the serial dependencies of FTS. To balance accuracy and efficiency, the optimization in \eqref{opt functional filter} identifies a small truncation lag $L_k$ such that
$$\sum_{|l|\leq L_k} \|\phi_{kl}\|^2 \geq 1 - \varepsilon_{L_k}, $$
where $0<\varepsilon_{L_k}<1$ is a pre-specified tolerance threshold. This guarantees that the retained filters capture the essential dependency structure while reducing redundant terms.

\subsection{Unified Functional Principal Component Analysis}\label{section_22}
In this subsection, we demonstrate that the optimization \eqref{opt functional filter} unifies the KL expansions \eqref{SFPCA fts} and \eqref{dyn kl}.
To this end, we define an equivalent relation of $\mathcal{A}_k$ induced by the shift-invariant property: $\forall \Phi_k,\Phi_k^{\prime}\in \mathcal{A}_k$,
$\Phi_k=\Phi_k^{\prime}$ iff $\Phi_k$ and $\Phi_k^{\prime}$ can be shifted to each other. A quotient space of $\mathcal{A}_k$ given this equivalent relation can be represented as 
\begin{align}
    \tilde{\mathcal{A}}_k:=\big\{\Phi_k\in \mathcal{A}_k;\max_{l\in \mathbb{Z}}\|\phi_{kl}\|= \|\phi_{k0}\|\big\},
    \label{quotient_space}
\end{align}
where the maximum is always obtainable by Proposition \ref{proposition2} (c). 
The above space indicates that we can always shift $\Phi_k\in\mathcal{A}_k$ such that its $L^{\infty}$ norm is $\|\phi_{k0}\|$.

\begin{theorem}[Solution to Optimal Functional Filters]
For stationary FTS $\{\varepsilon_j; j \in \mathbb{Z}\}$, the optimal functional filters defined in \eqref{opt functional filter} can be obtained by solving:
\begin{align}
\max_{\Phi_k\in\mathcal{A}_k}\|\Phi_k\|_{\infty}^2=\max_{\Phi_k\in\tilde{\mathcal{A}}_k}\|\Phi_k\|_{\infty}^2
    =\max_{\nu_k(\cdot)\in \mathcal{M}} \frac{1}{4 \pi^2}  \int_{-\pi}^{\pi} \int_{-\pi}^{\pi} \Psi_k(\omega_1, \omega_2) \overline{\nu_k(\omega_1)}\nu_k(\omega_2) \ \mathrm{d} \omega_1 \mathrm{d} \omega_2 , \label{simple opt functional filter}
\end{align}
where the kernels $\{\Psi_k(\cdot, \cdot); k \geq 1\}$ are defined as
\begin{align}
    \Psi_k(\omega_1, \omega_2) = \int_{0}^1\overline{\psi_k(t \mid \omega_1)}\psi_k(t \mid \omega_2)\ \mathrm{d} t, \quad k \geq 1, \  \omega_1, \omega_2 \in [-\pi, \pi],
    \label{psi kernel}
\end{align}
with $\psi_k(\cdot \mid \omega)$ being any valid $k$th eigenfunction of $f(\cdot,\cdot\mid \omega)$.
Accordingly, the optimal functional filters are constructed by $\tilde{\Phi}_k = \big\{\phi_{kl}(\cdot \mid \tilde{\nu}_k);l\in \mathbb{Z}\big\}$, where
\begin{align}
    \phi_{kl}(t \mid \tilde{\nu}_k)= \frac{1}{2\pi} \int_{-\pi}^{\pi}\psi_k(t \mid \omega)\tilde{\nu}_k(\omega) \exp (-\operatorname{i}l\omega) \mathrm{d} \omega, \quad t \in [0,1], \ l \in \mathbb{Z},
    \label{functional_filter_nu}
\end{align}
with $\tilde{\nu}_k(\cdot)$ being the maximizer of \eqref{simple opt functional filter}.
With this, the dynamic KL expansion for $\varepsilon_j$ in \eqref{dyn kl} is constructed as
\begin{align}
    \varepsilon_j(t)=\sum_{k=1}^{\infty}\sum_{l \in \mathbb{Z}} \phi_{kl}(t \mid \tilde{\nu}_k)\xi_{(j+l)k} \quad
        \text{with} \quad \xi_{jk} = \sum_{l \in \mathbb{Z}} \langle \varepsilon_{j-l}, \phi_{kl}(\cdot \mid \tilde{\nu}_k)\rangle, \quad t \in [0,1].
        \label{opt dyn kl}
\end{align}
\label{Theo_functional_filter}
\end{theorem}

Proof of Theorem \ref{Theo_functional_filter} can be found in the Supplementary Materials. In Theorem \ref{Theo_functional_filter}, although the kernel $\Psi_k(\cdot, \cdot)$ is not unique, the collection of optimal functional filters $\tilde{\Phi}_k$ remains the same for any valid eigenfunction $\psi_k(\cdot \mid \omega)$; see the Supplementary Materials for details. By this theorem, we transform the non-trivial optimization problem \eqref{opt functional filter} into a doable constrained maximization problem \eqref{simple opt functional filter}. 
It is worth noting that $\Psi_k(\cdot, \cdot)$ is Hermitian, i.e., $\Psi_k(\omega_1, \omega_2) = \overline{\Psi_k(\omega_2, \omega_1)}$ for all $\omega_1, \omega_2 \in [-\pi, \pi]$. As such, the optimization problem \eqref{simple opt functional filter} reduces to finding the eigenfunction of the hermitian kernel \( \Psi_k(\cdot, \cdot) \) corresponding to its largest eigenvalue, where the eigenfunction \( \tilde{\nu}_k(\cdot) \) is valued in \( \mathcal{M} \) rather than being normalized in the usual \( L^2 \) sense. By the definition of \( \mathcal{M} \), both \( \tilde{\nu}_k(\cdot) \) and \( -\tilde{\nu}_k(\cdot) \) are maximizers of \eqref{simple opt functional filter}.

Theorem \ref{Theo_functional_filter} is closely related to the concept of weak separability in functional data analysis \citep{lynch2018test, liang2022test, zapata2022partial, tan2024graphical}, which defines a specific form of serial dependence structure of FTS through the dependencies of scores in \eqref{SFPCA fts}. The definition is given as follows:

\begin{definition}[Serial Weak Separability]\label{def_1}
We say that a stationary FTS $\{\varepsilon_j; j \in \mathbb{Z} \}$ is weakly separable if there exist orthonormal basis functions of $L^2\left([0,1], \mathbb{R}\right)$, denoted as $\left\{\varphi^0_k(\cdot); k \geq 1\right\}$, such that $\operatorname{cov}(\xi^0_{j_1k_1}, \xi^0_{j_2k_2})=0$ for any $j_1, j_2 \in \mathbb{Z}$ whenever $k_1 \neq k_2$, where $\xi^0_{jk} = \langle \varepsilon_{j}, \varphi^0_{k}\rangle$. \label{def1}
\end{definition}

The above definition is weaker than the separability condition for FTS that assumes $c_g(t,s) = c^{(1)}(g) \cdot c^{(2)}(t,s)$,
which decomposes the auto-covariance into terms $c^{(1)}(g)$ and $c^{(2)}(t,s)$, characterizing the dependencies in the discrete time domain $j \in \mathbb{Z}$ and the functional domain $t \in [0,1]$, respectively.
We refer to Definition~\ref{def1} as \textbf{serial weak separability} for $\{\varepsilon_j; j \in \mathbb{Z} \}$ in what follows. It is straightforward to show that when $\{\varepsilon_j; j \in \mathbb{Z} \}$ satisfies the serial weak separability, $\varphi^0_k(\cdot)$ can be chosen as $\varphi_k(\cdot)$ in \eqref{SFPCA fts} for all $k$, and the resulting FPC scores satisfy $\operatorname{cov}(\xi_{j_1k_1}, \xi_{j_2k_2}) = 0$ for any $j_1$ and $j_2$ whenever $k_1 \neq k_2$.
In other words, the serial weak separability of $\{\varepsilon_j; j \in \mathbb{Z} \}$ allows modeling the scores $\{\xi_{jk}; j \in \mathbb{Z}\}$ without considering cross-correlations among different components. Similar conditions have been adopted for FTS in literature \citep{hyndman2007robust, hyndman2009forecasting}.

\begin{theorem}[Equivalent Conditions of Serial Weak Separability]
For a stationary FTS $\{\varepsilon_j; j \in \mathbb{Z} \}$, the following four statements are equivalent:
\begin{itemize}
    \item[\textnormal{(a)}] The serial weak separability is achieved.

    \item[\textnormal{(b)}] The auto-covariances $c_{h}(\cdot,\cdot)$ can be represented as
    \begin{align}
    c_{h}(t,s) = \sum_{k = 1}^{\infty} \lambda_{hk} \varphi_{k}(t) \varphi_{k}(s), \quad t,s \in [0,1], \ h \in \mathbb{Z},
    \label{autocov_structure}
\end{align}
where $\lambda_{hk}$'s and $\varphi_k(\cdot)$'s are the eigenvalues and eigenfunctions of $c_{h}(\cdot,\cdot)$ for each $h$.

    \item[\textnormal{(c)}] The eigenfunctions $\{\psi_k(\cdot \mid \omega); k \geq 1\}$ can be separated as 
    $\psi_k(t \mid \omega) = \gamma_k(\omega)\varphi_k(t)$ for $ t \in [0,1]$ and $\omega \in [-\pi, \pi]$,
     where $\gamma_k(\cdot)\in \mathcal{M}$, and the spectral density function $f(\cdot, \cdot \mid \omega)$ can be represented as 
     \begin{align*}
         f(t, s \mid \omega)=\sum_{k=1}^{\infty} \eta_k(\omega) \varphi_{k}(t) \varphi_{k}(s),  \quad t,s \in [0,1], \ \omega \in [-\pi, \pi].
     \end{align*}
     \item[\textnormal{(d)}] The kernel $\Psi_k(\cdot, \cdot)$ defined in Theorem \ref{Theo_functional_filter} can be decomposed as
\begin{align}\label{dec_Psi}
    \Psi_k(\omega_1, \omega_2) = \overline{\gamma_k(\omega_1)}\gamma_k(\omega_2), \quad \omega_1,\omega_2 \in [-\pi, \pi],
\end{align}
where $\gamma_k(\cdot)$ is contained in $\mathcal{M}$. Accordingly, $\max_{\Phi_k\in\mathcal{A}_k}\|\Phi_k\|_{\infty}^2=1$, $\forall k\geq 1$. 
\end{itemize}
\label{equal FPCA}
\end{theorem}

In Theorem~\ref{equal FPCA}, the equivalence of (a) and (b) has been proven by \citet{liang2022test}, which means that all the lag-$h$ covariance functions share the same eigenfunctions if the weak separability holds. In addition, part (c) shows that $f(\cdot, \cdot | \omega)$ at any frequency $\omega \in [-\pi,\pi]$ also shares the same set of eigenfunctions $\{\varphi_k(\cdot); k \geq 1\}$ under the serial weak separability condition. This leads to (d) by applying Theorem \ref{Theo_functional_filter}. Moreover, by using the decomposition \eqref{dec_Psi}, the optimal functional filters can be obtained by setting $\tilde{\nu}_k(\cdot)$ in \eqref{functional_filter_nu} as $\overline{\tilde{\nu}_k(\omega)}$, according to Theorem~\ref{Theo_functional_filter}. Then we have
$$  
\phi_{kl}(t \mid \tilde{\nu}_k) = \left\{
\begin{array}{lcr}
 \frac{1}{2\pi} \int_{-\pi}^{\pi} \varphi_k(t)\tilde{\nu}_k(\omega)\overline{\tilde{\nu}_k(\omega)}\exp (-\operatorname{i} l \omega) \mathrm{d} \omega = \varphi_k(t)& & l=0 \\
  0 &  & l \neq 0 
\end{array}
\right.
$$
under the serial weak separability. Detailed proof of Theorem~\ref{equal FPCA} can be found in the Supplementary Material.

\paragraph*{Remark} Theorem \ref{Theo_functional_filter} and \ref{equal FPCA} demonstrate that optimal functional filters unify diverse FPCA methodologies under a cohesive framework. When serial weak separability holds, eigenfunctions $\{\varphi_k(\cdot); k \geq 1\}$ derived from distinct covariance structures---such as the lag-0 covariance function $c_0(t_1,t_2)$ \citep{yao2005functional, li2010uniform}, aggregated lag-$h$ auto-covariance  $\sum_{h \in \mathbb{Z}} c_h(t_1,t_2)$ \citep{gao2019high}, auto-covariance integrals $\sum_{h \neq 0} \int_0^1 c_h(t,t_1)c_h(t,t_2)\ \mathrm{d}t$ \citep{bathia2010identifying} or the spectral density kernel $f(t, s \mid \omega)$ \citep{hormann2015dynamic}---coincide with those from the static Karhunen-Loève expansion \eqref{SFPCA fts}. Conversely, when serial weak separability fails, the dynamic KL expansion  \eqref{dyn kl} becomes the theoretically optimal representation for FTS, with Theorem \ref{Theo_functional_filter} ensuring its maximal parsimony. More importantly, our framework autonomously adapts to the underlying dependency structure without requiring prior validation of weak separability, achieving both theoretical generality and practical flexibility through dependency-adaptive dimension reduction.

\section{Estimation}\label{section3}

In this section, we assume that the latent functional time series $\{X_j(t);j\in \mathbb{Z}\}$ are contaminated, and we only have discrete observations
\begin{align}
Y_{jz}=X_j(t_{jz})+\tau_{j}(t_{jz})=\mu\left(t_{jz}\right)+\varepsilon_{ j}\left(t_{jz}\right)+\tau_{j}(t_{jz}),\quad  j=1, \ldots, J,\ z=1, \ldots, N_{j}, \label{obs model} 
\end{align}
where $\{t_{jz}\in [0,1]\}$ are observation times for the $j$th curve, $\tau_{j}(t_z)\sim N(0,\sigma^2)$ represents independent Gaussian noise, $\mu(\cdot)$ is the mean function of $X_j(\cdot)$'s. We further assume that  $\big\{t_{jz}\big\}$ and their total number $N_j$ vary across $j$, reflecting practical data constraints: sparse and/or irregular samples. Under the finite sample setting, we approximate the zero-mean process $\varepsilon_j(t):=X_j(t)-\mu(t)$ using a finite truncation
\begin{align}
\varepsilon_{ j}\left(t\right)\approx \sum_{k=1}^K\sum_{|l|\leq L_k}\phi_{kl}(t \mid \tilde{\nu}_k)\xi_{(j+l)k},\quad j= 1,\dots,J,\ t\in [0,1]\label{latent model},
\end{align}
where $K$ and $L_k$ are truncation numbers, $\phi_{kl}(\cdot \mid \tilde{\nu}_k)$ are the optimal functional filters given by \eqref{functional_filter_nu}. For each $k$, $\{\xi_{jk};j\in \mathbb{Z}\}$ is a mean-zero stationary scalar time series with spectral density $\eta_k(\omega)$, $\omega \in [-\pi, \pi]$.

We emphasize that the expression in \eqref{latent model} is different from the dynamic KL expansion proposed by \citet{hormann2015dynamic}. First, we employ the optimal functional filters \( \phi_{kl}(\cdot \mid \tilde{\nu}_k) \) instead of arbitrary functional filters, obtaining a more parsimonious representation and including the static KL expansion as a special case when serial weak separability holds. Second, we do not assume that the dynamic FPC scores can be derived by some projection method such as 
\begin{equation}\label{projection}
  \xi_{jk} = \sum_{l \in \mathbb{Z}} \langle \varepsilon_{j-l}, \phi_{kl}\rangle.
\end{equation}
We raise this issue because the projection method estimates $\xi_{jk}$ at boundaries using unobserved future or past curves $\big\{\varepsilon_{j}(\cdot);j>J\ \text{or}\ j<1\}$, which are usually set to zero in practice. This would introduce estimation biases, as discussed in Section 4 of \citet{hormann2015dynamic}. Alternatively, our approach relieves boundary biases by using the Whittle likelihood \citep{whittle1961gaussian} within a Bayesian framework. This adaptation ensures that estimation and prediction remain feasible even with finite, irregularly sampled data.

\subsection{Estimation of Mean Function and Optimal Functional Filters}\label{section3_1}

We estimate the mean function of $X_j(\cdot)$'s by using a local linear smoother, 
\begin{align}
   \underset{(a_0, a_1) \in \mathbb{R}^2}{\arg \min } \frac{1}{J} \sum_{j=1}^J \frac{1}{N_j} \sum_{z=1}^{N_j} K_{B_\mu}\big(t_{jz}-t\big)\cdot\big\{Y_{jz}-a_0 - a_1\left(t_{jz}-t\right)\big\}^2,
   \label{mean_est}
\end{align}
where $K_{B_\mu}(\cdot)$ is an univariate kernel function with bandwidth $B_\mu > 0$, and $a_0$ and $a_1$ are coefficients to be determined. Denote the minimizer $a_0$ as $\hat{\mu}(t)$, it serves as the local linear estimator of $\mu(t)$. For more details, refer to Chapter 8 of \citet{hsing2015theoretical}.

Similarly, we estimate the spectral density kernel $f(\cdot,\cdot|\omega)$, for any $\omega\in [-\pi,\pi]$, using a local surface smoother \citep{rubin2020sparsely}. Specifically, let 
$$
c_h(t_{(j+h)z_1}, t_{jz_2})=\big\{Y_{(j+h)z_1}-{\mu}(t_{(j+h) z_1})\big\}\cdot\big\{Y_{jz_2}-{\mu}(t_{j z_2})\big\},
$$
and denote its unbiased estimator as $\hat{c}_{hj}\big(t_{(j+h)z_1}, t_{jz_2}\big)$, where the mean functions are replaced by their estimator from \eqref{mean_est}.
We then consider the minimization
\begin{align}
\begin{aligned}
\mathop{\arg \min} _{\left(d_0, d_1, d_2\right) \in \mathbb{C}^3} 
 &\frac{1}{L} \sum_{h=-L}^L  \frac{W_h}{J-|h|} \sum_{j=\max (1,1-h)}^{\min (J, J-h)} \frac{1}{M_{jh}} \sum_{1\leq z_1\leq N_{j+h}, 1\leq z_2\leq N_j}^{z_1\neq z_2\ \text{if}\ h=0} \bigg\{\hat{c}_{h j}\left(t_{(j+h) z_1}, t_{j z_2}\right) 
 \cdot \exp(\mathrm{i} h \omega) \\
 & -d_0 -d_1\left(t_{(j+h) z_1}-t\right)-d_2\left(t_{j z_2}-s\right)\bigg\}^2 \cdot K_{B_f}\left(t_{(j+h) z_1}-t\right) K_{B_f}\left(t_{j z_2}-s\right),
\end{aligned}
\label{spectral_est}
\end{align}
where $B_f > 0$ is a bandwidth, $L$ is the truncation of time lags, $W_h$s are positive weights for each time lag $h$, and $M_{jh}$ is defined as $N_{j+h} \cdot N_{j}$ if $h \neq 0$ and $N_{j} \cdot (N_{j}-1)$ if $h = 0$.
For any given $t,s$ and $\omega$, we denote the minimizer  $d_0$ of \eqref{spectral_est} as $\hat{d}_{B_{f}}(t,s \mid \omega)$. Then, $\hat{d}_{B_{f}}(t,s \mid \omega)$ is an estimator for 
\begin{align*}
  \frac{1}{\sum_{|h|\leq L}W_h}\sum_{|h|\leq L} W_h c_h(t,s)\exp(\mathrm{i} h \omega). 
\end{align*}
It is worth noting that $ \frac{1}{2\pi} \sum_{|h|\leq L} W_h c_h(t,s)\exp(\mathrm{i} h \omega) $ is the lag window estimator for $f(\cdot,\cdot|\omega)$ as proposed in \citet{hormann2015dynamic}. By adopting the Bartlett window, i.e., $W_h = (1 - |h|/L)$ for $|h| < L$, we then estimate $f(t,s|\omega)$ by
\begin{align}
    \hat{f}(t,s \mid \omega) =\frac{\sum_{|h|\leq L}W_h}{2\pi}\hat{d}_{B_{f}}(t,s \mid \omega)= \frac{L}{2\pi}\hat{d}_{B_{f}}(t,s \mid \omega).
    \label{Bartlett_est}
\end{align}
The selection rules of $B_{\mu}$, $B_{f}$, and $L$ are discussed in Supplementary Materials.

Now we perform a spectral decomposition on $\hat{f}(\cdot,\cdot \mid \omega)$ to estimate the eigenfunctions $\psi_{k}(\cdot \mid \omega)$. This can be done by computing the matrix eigendecomposition of $\hat{f}(\cdot,\cdot \mid \omega)$ on some dense discrete time grids, yielding the estimator $\hat{\psi}_{k}(\cdot \mid \omega)$. Given that, we estimate the kernel $\Psi_k(\cdot, \cdot)$ in \eqref{psi kernel} by 
\begin{align}
    \hat{\Psi}_k(\omega_1, \omega_2) = \int_{0}^1\overline{\hat{\psi}_k(t \mid \omega_{1})}\hat{\psi}_k(t \mid \omega_{2})\ \mathrm{d} t, \quad \omega_{1}, \omega_{2} \in [-\pi,\pi],
    \label{psi kernel est}
\end{align}
and solve the following optimization
\begin{align}
    \hat{\nu}_k(\cdot):=\mathop{\arg \min}_{\nu_k(\cdot)\in \mathcal{M}}\int_{-\pi}^{\pi} \int_{-\pi}^{\pi} \hat{\Psi}_k(\omega_1, \omega_2) \overline{\nu_k(\omega_1)}\nu_k(\omega_2) \ \mathrm{d} \omega_1 \mathrm{d} \omega_2,
    \label{simple opt functional filter est}
\end{align}
to estimate optimal functional filters. 

For simplification, we employ a discrete approximation for \eqref{simple opt functional filter est} and instead solve: 
\begin{align}\label{optff}
    \mathop{\arg\min}\limits_{\boldsymbol{\nu}_k \in \mathcal{M}_{s}} \mathcal{L}(\boldsymbol{\nu}_k) := \boldsymbol{\nu}_k^* \hat{\boldsymbol{\Psi}}_k \boldsymbol{\nu}_k,
\end{align}
where 
$
\boldsymbol{\nu}_k = (\nu_{k}(\omega_{-s}), \dots, \nu_{k}(\omega_s))^{*}, \quad \hat{\boldsymbol{\Psi}}_k = (\hat{\Psi}_k(\omega_l, \omega_m))_{-s \leq l,m \leq s},
$
and 
$
\mathcal{M}_{s} = \big\{(\nu_{-s}, \ldots, \nu_s); \nu_i \in \mathbb{C}, |\nu_i| = 1, \nu_i = \overline{\nu_{-i}}, i = -s, \ldots, s \big\},
$
with $ \mathcal{S}:= \{\omega_{-s}, \dots, \omega_s\}$ being an equally spaced dense subset of $[-\pi, \pi]$. We adopt a projected gradient method \citep{hastie2015statistical} to iteratively solve the constrained maximization problem \eqref{optff}. After obtaining $\hat{\boldsymbol{\nu}}_k$, we perform a numerical integration to calculate the optimal functional filters:
\begin{align}\label{est_functional_filter}
\hat{\phi}_{kl}(t \mid \hat{\nu}_k)= \frac{1}{2\pi} \int_{-\pi}^{\pi} \hat{\psi}_k(t \mid \omega)\hat{\nu}_k(\omega) \exp (-\operatorname{i}l\omega) \mathrm{d} \omega, \quad t \in [0,1], \ l \in \mathbb{Z}.
\end{align}
The details of the above procedure are summarized in Algorithm~\ref{algo1}.

\normalem
\begin{algorithm}[!ht]
\footnotesize
        \SetKwInOut{Input}{Input}
        \SetKwInOut{Output}{Output}
        \SetKwRepeat{Repeat}{Repeat}{Until}
        \SetKwFor{For}{For}{do}

\textbf{Input}: Initial vector ${{\boldsymbol{\nu}}}_k^{(1)}$, matrix $\hat{\boldsymbol{\Psi}}_k$, frequency set $\mathcal{S}$, estimated eigenfunction $\hat{\psi}_k(\cdot \mid \omega)$, pre-set threshold $\varepsilon_{L_k}$.
\\  $i = 1$; \\

\Repeat{\textnormal{the sequence $\{\mathcal{L}({\boldsymbol{\nu}}_k^{(i)});i\geq 1\}$ converges}}{
  $\boldsymbol{\tilde{\nu}}^{(i+1)}_k = {\boldsymbol{\nu}}^{(i)}_k + {\alpha}
\cdot (\hat{\boldsymbol{\Psi}}_k+\hat{\boldsymbol{\Psi}}_k^{*}){\boldsymbol{\nu}}^{(i)}_k$, with a step size ${\alpha}$ selected by the limited minimization rule \citep{hastie2015statistical}} 

\For{$\omega \in \mathcal{S}$}{
${\boldsymbol{\nu}}^{(i+1)}_k(\omega) = \boldsymbol{\tilde{\nu}}^{(i+1)}_k(\omega)/\big|\boldsymbol{\tilde{\nu}}^{(i+1)}_k(\omega)\big|$;
}
  $i = i + 1$; \\
Let $\hat{\nu}_k(\omega)$ be the element in $\tilde{\boldsymbol{\nu}}_k^{(i)}$ corresponding to frequency $\omega$, and set $L_k=0$;\\

\Repeat{$\sum_{|l|\leq L_k}\|\hat{\phi}_{kl}(\cdot \mid \hat{\nu}_k)\|^2 \geq 1 - \varepsilon_{L_k}$}{
$\hat{\phi}_{kl}(t \mid \hat{\nu}_k)= \frac{1}{s} \sum_{\omega \in \mathcal{S}} \hat{\psi}_k(t \mid \omega){\hat{\nu}}_k(\omega) \exp (-\operatorname{i}l\omega)  , \quad l = -L_k\ \text{and}\ L_k,\ t \in [0,1]$;   \\
$L_k=L_k+1$;\\
}
$\hat{\phi}_{kl}(t \mid \hat{\nu}_k)=\hat{\phi}_{kl}(t \mid \hat{\nu}_k)/\sum_{|l|\leq L_k}\|\hat{\phi}_{kl}(\cdot \mid \hat{\nu}_k)\|^2$, $\quad |l| \leq L_k,\ t \in [0,1]$;\\
\textbf{Output}: $\{\hat{\phi}_{kl}(\cdot \mid \hat{\nu}_k);|l|\leq L_k\}$.
\caption{{Projected Gradient Method for Optimal Functional Filters}}
\label{algo1}
\end{algorithm}
\ULforem

\paragraph*{Remark} In Algorithm~\ref{algo1}, we determine \(L_k\) by finding the minimal value such that 
$
\sum_{|l| \leq L_k} \|\hat{\phi}_{kl}(\cdot \mid \hat{\nu}_k)\|^2 \geq 1 - \varepsilon_{L_k},
$
as in \citet{hormann2015dynamic}, where \(\varepsilon_{L_k}\) is some pre-specified small value. It should be noted that the output of $L_k$ can be 0, which indicates that only a single filter is selected. By Theorem 2, this means that the serial weak separability condition holds for the data, and the optimal filter coincides with its corresponding principal component in static FPCA.  More importantly, this Algorithm autonomously determines whether to use dynamic FPCA.

\subsection{Bayesian Procedures for FPC Scores}\label{section_32}
Denote $\boldsymbol{\xi}_k := (\xi_{(1-L_k)k}, \ldots, \xi_{(J+L_k)k})$ as the dynamic FPC scores corresponding to the $k$th optimal functional filters. We model $\boldsymbol{\xi}_k$ using a Bayesian framework. Note that the joint posterior distribution of $\boldsymbol{\xi}_1, \ldots, \boldsymbol{\xi}_K$ given data can be written as
\begin{align}
    \pi \left(\boldsymbol{\xi}_1, \ldots, \boldsymbol{\xi}_K \mid \boldsymbol{Y} \right) \propto \mathcal{L}(\boldsymbol{\xi}_1, \ldots, \boldsymbol{\xi}_K \mid \boldsymbol{Y}) \cdot  \prod_{k=1}^K \pi (\boldsymbol{\xi}_k) , \label{posterior}
\end{align}
where $\pi(a\mid \cdot)\propto b$ means that $\pi(a\mid \cdot)=c\cdot b$ with $c$ being a constant independent of $a$,  $\boldsymbol{Y} = \{Y_{jz}; j = 1,\ldots, J, z = 1,\ldots, N_j\}$, and $\pi(\cdot)$ represents the prior distribution. Under Gaussianity assumption, the likelihood function $\mathcal{L}(\boldsymbol{\xi}_1, \ldots, \boldsymbol{\xi}_K \mid \boldsymbol{Y})$ is given as
\begin{align*}
    \mathcal{L}(\boldsymbol{\xi}_1, \ldots, \boldsymbol{\xi}_K \mid \boldsymbol{Y}) \propto \exp{\left[- \sum_{j=1}^J \sum_{z=1}^{N_{j}} \frac{\left\{Y_{j z}-\mu\left(t_{jz}\right)-\sum_{k \leq K} \sum_{|l| \leq L_k} \phi_{kl}(t_{jz} \mid \tilde{\nu}_k) \xi_{(j+l) k}\right\}^2}{2 \sigma^2}\right] },
\end{align*}
where $\mu(\cdot)$ and $\phi_{kl}(\cdot \mid \tilde{\nu}_k)$ can be substituted by their respective estimates $\hat{\mu}(\cdot)$ and $\hat{\phi}_{kl}(\cdot \mid \hat{\nu}_k)$, and $\sigma^2$ can be estimated, for example, by the approach in \citet{yao2005functional} from observed data.

Notice that $\{\xi_{jk};j\in \mathbb{Z}\}$ is a weakly stationary time series with spectral density $\eta_k(\cdot)$, 
we adopt the Whittle likelihood \citep{whittle1961gaussian, subba2021reconciling} to construct a prior distribution for $\boldsymbol{\xi}_k$ in the frequency domain. To this end, let 
\begin{align*}
\boldsymbol{\tilde{\xi}}_k(\omega) = \frac{1}{\sqrt{2\pi(J+2L_k)}}\sum_{j = 1}^{J+2L_k}\xi_{(j-L_k)k}\exp(\mathrm{i} j \omega), \quad \omega \in \mathcal{S}_J, 
\end{align*}
be the discrete Fourier 
transformation of $\bm{\xi}_k$,  where $\mathcal{S}_J := \{\omega_j = \frac{2\pi j}{J}; j = 1,\ldots, J\}$. Asymptotic result suggests that $\boldsymbol{\tilde{\xi}}_k(\omega)$, $\omega \in \mathcal{S}_J$, behaves as independent complex mean-zero Gaussian variables with variances $\eta_k(\omega)$ as $J\rightarrow\infty$ \citep{whittle1961gaussian}.
Correspondingly, the log-likelihood of the prior can be approximated by
\begin{align}
    \log \pi (\boldsymbol{\xi}_k) \approx -\frac{1}{2} \sum\limits_{j=1}^{J}    \bigg[\frac{|\tilde{\boldsymbol{\xi}}_k\left(\omega_j\right)|^2 }{\eta_k\left(\omega_j\right)}+\log\left\{\eta_k\left(\omega_j\right)\right\}\bigg], \quad \omega_j \in \mathcal{S}_J, \label{whittle}
\end{align}
where $\eta_k(\omega)$ can be estimated by the eigenvalue of $\hat{f}(\cdot,\cdot \mid \omega)$ obtained from \eqref{Bartlett_est}.

We propose a gradient ascend algorithm to obtain the  Maximum A Posteriori (MAP) estimator of the dynamic FPC scores based on \eqref{posterior}. For convenience, we proceed with the log-transformed posterior distribution, where the gradient of the FPC scores is given by 
\begin{align*}
& \frac{\partial \log \pi\left(\boldsymbol{\xi}_1, \ldots, \boldsymbol{\xi}_K \mid \boldsymbol{Y} \right)}{\partial \boldsymbol{\xi}_k} \nonumber \\
=&  - {\sigma}^{-2}\left\{ \sum_{k^{\prime} = 1}^K \boldsymbol{\xi}_{k^{\prime}}\sum_{j=1}^{J}{\boldsymbol{\phi}}^{*}_{k^{\prime} j}{\boldsymbol{\phi}}_{k j} - \sum_{j=1}^{J} \tilde{\boldsymbol{Y}}_j{\boldsymbol{\phi}}_{kj}\right\} \nonumber  
- \operatorname{Re}\left(\boldsymbol{\xi}_k \sum_{j=1}^{J}\{{\eta}_k\left(\omega_j\right)\}^{-1}  \boldsymbol{\rho}_k\left(\omega_j\right) \{\boldsymbol{\rho}_k\left(\omega_j\right)\}^{*}\right),
\end{align*}
where $\tilde{\boldsymbol{Y}}_j:= (Y_{j1} - {\mu}(t_{j1}), \ldots, Y_{jN_j} - {\mu}(t_{jN_j}))$ denotes the demeaned observations, ${\boldsymbol{\phi}}_{k j}$ is a $N_j \times (J+2L_k)$ matrix with the $(z, j + l + L_k)$th element being $\phi_{kl}(t_{jz} \mid \tilde{\nu}_k)$ for $|l| \leq L_k$ and 0 otherwise, $\boldsymbol{\rho}_k\left(\omega\right):=\frac{1}{\sqrt{2\pi(J+2 L_k)}}\left(\exp \left(\mathrm{i} 1 \omega\right), \ldots, \exp \left(\mathrm{i}\left(J+2 L_k\right) \omega\right)\right)^{*}$, and $\operatorname{Re}(\cdot)$ is the operation to extract the real part of a complex matrix. Using the gradient above, we optimize the log-posterior distribution \( \log \pi\left(\boldsymbol{\xi}_1, \ldots, \boldsymbol{\xi}_K \mid \boldsymbol{Y} \right) \) w.r.t. the scores via gradient ascent. The resulting maximizer serves as the MAP estimator for the scores, denoted \( \hat{\boldsymbol{\xi}}_1, \ldots, \hat{\boldsymbol{\xi}}_K \).

\subsection{Curve Reconstruction and Forecast} \label{section_33}
Given the MAP estimator above, we reconstruct the underlying curves $X_j(\cdot)$ by
\begin{align}
    \hat{X}_j(t) = \hat{\mu}(t) + \sum_{k=1}^{K}\sum_{|l| \leq L_k} \hat{\phi}_{kl}(t \mid \hat{\nu}_k)\hat{\xi}_{(j+l)k}. \quad j = 1,\ldots, J, \ t \in [0,1]. \label{reconstruction}
\end{align}
The selection of $K$ is detailed in the Supplementary Materials.

\normalem
\begin{algorithm}[!ht]

\footnotesize
        \SetKwInOut{Input}{Input}
        \SetKwInOut{Output}{Output}
        \SetKwRepeat{Repeat}{Repeat}{Until}
        \SetKwFor{For}{For}{do}

\textbf{Input}: Estimated mean function $\hat{\mu}(\cdot)$, optimal functional filters $\{ \hat{\phi}_{kl}(\cdot \mid \hat{\nu}_k); |l| \leq L_k, k = 1,\ldots, K\}$, dynamic FPC scores $\{\hat{\boldsymbol{\xi}}_k; k = 1,\ldots, K \}$, and prediction operators $\mathcal{P}_k(\cdot;P)$, $k\leq K$, with $P$ being the prediction length.

\For{$k = 1, \ldots, K$}{
Predict the dynamic FPC scores for $P$ steps ahead as 
\begin{align*}
(\hat{\xi}_{(J+L_k+1)k}, \ldots, \hat{\xi}_{(J+L_k+P)k}) = \mathcal{P}_k(\hat{\boldsymbol{\xi}}_k;P);
\end{align*}
}

\For{$p = 1, \ldots, P$}{
Calculate the $p$-step forecast of FTS as
\begin{align}
\hat{X}_{J+p}(t) = \hat{\mu}(t) + \sum_{k=1}^{K}\sum_{|l| \leq L_k} \hat{\phi}_{kl}(t \mid \hat{\nu}_k)\hat{\xi}_{(J+p+l)k}, \quad t \in [0,1]; \label{FTS_prediction}
\end{align}
}

\textbf{Output}: FTS forecast $\{\hat{X}_{J+p}(\cdot);p = 1,\ldots,P\}$.

\caption{{FTS Forecast using PADA}}
\label{algo2}
\end{algorithm}
\ULforem

We employ the above reconstruction \eqref{reconstruction} for the forecasting of functional time series. Since dynamic FPC scores $\xi_{jk}$s are mutually uncorrelated for different $k$, each $\boldsymbol{\xi}_k$ is a univariate time series and can be forecast separately. The general algorithm is summarized in Algorithm \ref{algo2}, where $\mathcal{P}_k(\hat{\boldsymbol{\xi}}_k;P)$ is a $P$-step-ahead prediction operator. For the form of $\mathcal{P}_k$, we can use various standard time series models, such as AR models or ARMA models \citep{brillinger2001time, de200625}.

To measure uncertainties in reconstruction and forecast, we construct point-wise credible intervals of $X_j(\cdot)$ under the Bayesian framework. Given any specific $t$, the posterior distribution $\pi \left(X_j(t) \mid \boldsymbol{Y} \right)$ follows
\begin{align}
    \pi \left(X_j(t) \mid \boldsymbol{Y} \right)  \propto \int p\left(X_j(t)\mid \boldsymbol{\xi}_1, \ldots, \boldsymbol{\xi}_K \right) \pi \left( \boldsymbol{\xi}_1, \ldots, \boldsymbol{\xi}_K \mid \boldsymbol{Y} \right)  \ \mathrm{d}\boldsymbol{\xi}_1\dots \mathrm{d}\boldsymbol{\xi}_K,
    \label{pos_recon}
\end{align}
where $p\left(X_j(t)\mid \boldsymbol{\xi}_1, \ldots, \boldsymbol{\xi}_K\right)$ is a probability density determined by \eqref{latent model}, and $\pi \left( \boldsymbol{\xi}_1, \ldots, \boldsymbol{\xi}_K \mid \boldsymbol{Y} \right)$ is determined by \eqref{posterior}. Details about the form of \( \pi \left(X_j(t) \mid \boldsymbol{Y} \right) \) for \( j \leq J \) and \( j > J \), as well as how to compute the integral in \eqref{pos_recon}, can be found in the Supplementary Material.
Given \eqref{pos_recon}, we employ a Hamiltonian Monte Carlo algorithm \citep{thomas2021learning} to sample from $\pi \left(X_j(\cdot) \mid \boldsymbol{Y} \right)$, where the mean function, optimal functional filters, and the variance $\sigma^2$ are substituted by their estimates. Subsequently, the point-wise credible intervals of $X_j(t)$, $t\in [0,1]$, can be constructed using the quantiles of the posterior samples. Similarly, we  generate the posterior samples of $Y_j(t) := X_j(t) + \tau_j(t)$, $j>J$ and $t\in [0,1]$, for the point-wise credible interval of forecasts. Note that
\begin{align}
    \pi \left(Y_j(t) \mid \boldsymbol{Y} \right)  \propto \int p\left(Y_j(t)\mid X_j(\cdot)\right) \cdot \pi \left(X_j(\cdot) \mid \boldsymbol{Y} \right)\ \mathrm{d}X_j(\cdot),
    \label{pos_forecast}
\end{align}
where $p\left(Y_j(t)\mid X_j(\cdot)\right)$ is a Gaussian density determined by \eqref{obs model}. We use the posterior samples of $\pi \left(X_j(\cdot) \mid \boldsymbol{Y} \right)$ to generate samples from $\pi \left(Y_j(t) \mid \boldsymbol{Y} \right)$.

\section{Statistical Consistencies}\label{ST}
In this section, we investigate the statistical consistencies of optimal functional filters estimated from contaminated FTS data. To this end, we need to impose some identifiability conditions for eigenfunctions ${\psi}_k(\cdot \mid \omega)$ of the spectral density kernel $f(\cdot, \cdot \mid \omega)$ in \eqref{spectral decomp} and  $\tilde{\nu}_k(\cdot)$ in \eqref{simple opt functional filter}, respectively. For any given ${\psi}_k(\cdot \mid \omega)$, we follow \citet{hormann2015dynamic, tan2024graphical} and assume that $\langle {\psi}_k(\cdot \mid \omega), \hat{\psi}_k(\cdot \mid \omega) \rangle \geq 0$, where $\hat{\psi}_k(\cdot \mid \omega)$ is the eigenfunction of estimated spectral density kernel $\hat{f}(\cdot, \cdot \mid \omega)$ in \eqref{Bartlett_est}. This condition can be achieved, without loss of generality, by adjusting $\hat{\psi}_{k}(\cdot\mid\omega)$ to $\hat{\psi}_{k}(\cdot\mid\omega)\cdot \frac{|\langle\psi_{k}(\cdot\mid\omega),\hat{\psi}_{k}(\cdot\mid\omega)\rangle|}{\langle\psi_{k}(\cdot|\omega),\hat{\psi}_{k}(\cdot\mid\omega)\rangle}$ when $\langle\psi_{k}(\cdot\mid\omega),\hat{\psi}_{k}(\cdot\mid\omega)\rangle\neq 0$. 
Similarly, we assume that $\langle \tilde{\nu}_k, \hat{\nu}_k \rangle \geq 0$ for any given $\tilde{\nu}_k(\cdot)$, where $\hat{\nu}_k(\cdot)$ is defined in \eqref{simple opt functional filter est}.

In the following, we suppose that the random functions $X_j(\cdot)$ satisfy $\operatorname{E} | X_j(t) |^{s} < \infty$ for $j \in \mathbb{Z}$ and $t \in [0,1]$, for some $s>2$. In addition, $K_{B_{\mu}}(\cdot)$ and $K_{B_{f}}(\cdot)$ used in local smoothers \eqref{mean_est} and \eqref{spectral_est} are both in the form $K_{B}(u) = \frac{1}{B} K(u/B)$ for $u \in \mathbb{R}$, where $K(\cdot)$ is a kernel function on $\mathbb{R}$ and $B$ represents the bandwidth either $B_{\mu}$ or $B_f$.
In addition, we assume that the latent FTS $X_j(\cdot)$, the time grids $t_{jz}$, the numbers of time points $N_j$, and the measurement noises $\tau_{jz}$ are independent of each other. 

Based on the above settings, we impose the following assumptions.

\begin{assumption}[Strong mixing condition]
    $\{ X_j(\cdot); j \in \mathbb{Z} \}$ is a stationary and strong mixing sequence with the strong mixing coefficient $\alpha(h)$ defined by
    \begin{align*}
        \alpha(h) = \mathop{\sup}\limits_{A \in \mathcal{F}_{-\infty}^{0}, B \in \mathcal{F}_{h}^{\infty}} \big| \mathbb{P}(A \cap B) - \mathbb{P}(A)\mathbb{P}(B) \big|,
    \end{align*}
    where $\mathcal{F}_{j_1}^{j_2}$ denotes the $\sigma$-field generated by $\{ X_j(\cdot); j_1 \leq j \leq j_2 \}$. Here, $\alpha(h) \leq C h^{-\beta}$ with $C < \infty$, where $\beta$ satisfies $\beta > \frac{2s - 2}{s-2}$.
    \label{strong_mixing}
\end{assumption}

\begin{assumption}[Conditions on observation schemes]
The probability density for the random variables $t_{jz}$‘s is bounded away from 0 and continuous on $[0,1]$. Additionally, the numbers of time points $N_j$s are i.i.d. random variables valued in $\{1, \ldots, N_{max}\}$ with $N_{max} < \infty$. 
\label{obs_structure}
\end{assumption}

\begin{assumption}[Conditions on measurement errors]
The measurement errors $\tau_{jz}$s are i.i.d. random variables with zero mean, satisfying $\operatorname{E} |\tau_{jz}|^s < \infty$ for $s>2$. 
\label{measure_error_condition}
\end{assumption}

\begin{assumption}[Conditions on kernels]
    For an integrable and bounded kernel $K(\cdot)$, it is either a Lipschitz continuous function with compact support,
    or it has a bounded derivative, $| \frac{\partial}{\partial u} K(u) | \leq C$, and for some $\alpha > 1$ and $U_s < \infty$, $| \frac{\partial}{\partial u} K(u) | \leq C | u |^{-\alpha}$, for $|u| > U_s$.
    \label{kernel_assumption}
\end{assumption}

\begin{assumption}[Uniqueness and finite conditions on $\tilde{\nu}_k(\cdot)$]\label{uni_sparse_con}
For each \( 1 \leq k \leq K \), there exists a unique \( \tilde{\nu}_k(\cdot) \in \mathcal{M} \) such that the optimal solution to \eqref{simple opt functional filter} is \( \{ \tilde{\nu}_k(\cdot), -\tilde{\nu}_k(\cdot) \} \). In addition, 
\( \phi_{kl}(t \mid \tilde{\nu}_k) = 0 \) for \( t \in [0,1] \) and \( |l| > L_k \), where \( L_k \) is a non-negative integer.
\end{assumption}

Assumption \ref{strong_mixing} is a widely used condition to impose a short-term dependency on FTS data \citep{bosq2000linear, hansen2008uniform, rubin2020sparsely}.
This condition quantifies the degree of serial dependence using the strong mixing coefficient $\alpha(h)$, assuming that $\alpha(h)$ converges at a polynomial rate $h^{-\beta}$ as the time lag $h$ goes to infinity.  Assumptions \ref{obs_structure} - \ref{kernel_assumption}  introduce conditions on the observed time grids, measurement noises, and the kernel used in local smoothers. These assumptions are commonly adopted for sparsely and irregularly observed functional data \citep{yao2005functional, li2010uniform, wang2016functional,tan2024functional}. Assumption \ref{uni_sparse_con} indicates that the optimal functional filters for the first $K$ components are unique. Furthermore, it also states that the dynamic KL expansion of $\varepsilon_j(\cdot)$ with the first $K$ optimal functional filters is given by 
$$
\sum_{k=1}^{K}\sum_{|l| \leq L_k} \phi_{kl}(\cdot \mid \tilde{\nu}_k)\xi_{(j+l)k}.
$$
This assumption holds with $L_k=0$ if the serial weak separability in Definition \ref{def1} is satisfied; see Theorem \ref{equal FPCA}. For general cases, the condition ensures that the dynamic KL expansion with optimal functional filters has a finite representation for the first $K$ components.

In addition to the above assumptions, other regularity conditions are provided in the Supplementary Material.

For an estimated functional filter defined as \eqref{est_functional_filter}, we define the corresponding true functional filter as $\{{\phi}_{kl}(\cdot);l\in \mathbb{Z}\}$, which is calculated by substituting $\hat{\psi}_k(\cdot \mid \omega)$ in \eqref{est_functional_filter} with the true eigenfunction ${\psi}_k(\cdot \mid \omega)$ that satifies $\langle {\psi}_k(\cdot \mid \omega), \hat{\psi}_k(\cdot \mid \omega) \rangle \geq 0$. Then we have the statistical theories of optimal functional filters as follows.

\begin{theorem}\label{opt_rate}
Given Assumptions \ref{strong_mixing} - \ref{uni_sparse_con} and other regular conditions in the Supplementary Material, we further assume that for any $k_1, k_2 \leq K$, 
\begin{align}
    \min_{\omega \in [-\pi,\pi], k_1 \neq k_2} | \eta_{k_1}(\omega) - \eta_{k_2}(\omega) | > 0,
    \label{ident_con}
\end{align}
where $\eta_{k}(\omega)$ is the $k$th eigenvalue of $f(\cdot, \cdot \mid \omega)$ and $K$ is a finite number. Then for $k\leq K$, 
\begin{align}
\mathop{\sup}\limits_{\substack{l \in \mathbb{Z}}} \big\| \hat{\phi}_{kl}(\cdot \mid \hat{\nu}_k) - \phi_{kl} \big\| &= \mathcal{O}_{p} \bigg( L \sqrt{\frac{\log J}{J B^2_{f}}} + L B^2_{f}  \bigg),\label{ori_converge_rate}
\end{align}
with $\mathop{\sup}\limits_{\substack{|l| > L_k}} \big\| \hat{\phi}_{kl}(\cdot \mid \hat{\nu}_k)\big\|$ converging to 0 in probability, where $\{\phi_{kl} (\cdot);l\in \mathbb{Z}\}$ is the true functional filters corresponding to $\{\hat{\phi}_{kl}(\cdot \mid \hat{\nu}_k);l\in \mathbb{Z}\}$. Here, $L\rightarrow\infty$ and $B_f\rightarrow 0$ such that $\frac{\log J}{J^{\theta}B^2_{f}} = o(1)$, with $\theta = \frac{\beta \cdot (s-2) - 4s + 4}{\beta \cdot (s-2)}$, $B_{f}^2 = o(1/L)$, and $L = o\big(\big(\sqrt{\frac{\log J}{JB_{f}^2}}\big)^{-\frac{s-2}{s-1}}\big)$. 
\end{theorem}

Condition \eqref{ident_con} is a common assumption in the literature \citep{hormann2010weakly, hormann2015dynamic, tan2024graphical} to ensure the identifiability of eigenfunctions.
Under this condition, \( \hat{\phi}_{kl}(\cdot \mid \hat{\nu}_k) \) consistently estimates the true functional filters \( \phi_{kl}(\cdot) \) with convergence rate \( L \sqrt{\frac{\log J}{J B^2_{f}}} + L B^2_{f} \). 
This rate has a similar order to the eigenfunctions estimated from sparsely observed functional data \citep{li2010uniform, rubin2020sparsely}, distinct from the rate of functional filters estimated from densely observed functional data \citep{hormann2015dynamic,tan2024graphical}.

Moreover, Theorem \ref{opt_rate} indicates that $\{ \hat{\phi}_{kl}(\cdot \mid \hat{\nu}_k); |l| > L_k \}$ converges to 0 as $J \rightarrow \infty$. These terms represent the redundant parts in KL expansions under the Assumption \ref{uni_sparse_con}, and their convergence to 0 ensures the statistical consistency of PADA to the most parsimonious dynamic KL expansion.
In contrast, the functional filters proposed by \citet{hormann2015dynamic} cannot guarantee the convergence of the redundant parts as $J \rightarrow \infty$, which may result in redundant dynamic KL expansions that hinder its utility for predicting FTS.

\section{Simulation}\label{section4}

\subsection{Set-up and Data Generation}\label{section4_1}
In this section, we compare the proposed PADA with several existing methods in both curve reconstruction and forecast of FTS. We consider zero-mean FTS $\{\varepsilon_j(\cdot);j =1,\dots,J\}$ admitting the dynamic KL expansion:
\begin{align}
    \varepsilon_j(t)= \sum_{k = 1}^{K} \sum_{l = -L_{k,1} }^{L_{k,2}} w_{l} \phi_{kl}(t) \xi_{(j+l)k}, \quad j  = 1 ,\ldots ,J,\ t \in [0,1],
    \label{data gen}
\end{align}
where $L_{k,1}$s and $L_{k,2}$s are integers, $\omega_l$s denote positive weights, and $\phi_{kl}(\cdot)$ represent a collection of Fourier basis functions. To generate FTS data, we assume that $w_{l}=\sqrt{ w_{l}^{\prime} / \sum_{l = -L_{k,1}}^{L_{k,2}} w_{l}^{\prime}}$ with $w_{l}^{\prime}=\exp (-|l|/2)$ for each $k$. Under this setting, the weights satisfy $\sum_{l = -L_{k,1}}^{L_{k,2}} w_l^2=1$, and a smaller weight is assigned for $\phi_{kl}(\cdot)$ when $|l|$ increases. 
It can be shown that $\{w_{l} \phi_{kl}(\cdot);l\in \mathbb{Z}\}$ is the optimal functional filters with its $L^{\infty}$ norm as $w_0$ for each $k$; see the Supplementary Material for more details.
To introduce serial dependencies, we generate $\{\xi_{jk}, j=1,\ldots,J\}$ using an AR(1) model, i.e. $\xi_{(j+1)k}=\rho \xi_{jk}+b_{jk}$ for each $k$, where $\rho$ is taken as $0.2$ and $\{b_{jk},j\in \mathbb{Z}\}$ are i.i.d. Gaussian noise following $\mathcal{N}(0,1/k)$. The data generation is independent across different $k$.

Based on \eqref{data gen}, we generate the discrete observations of FTS by
\begin{align}\label{data measure}
    Y_{jz} = \varepsilon_j(t_{jz}) + \tau_{jz}, \quad j = 1,\ldots, J, \ z = 1,\ldots, N_j.
\end{align}
Specifically, we first construct an evenly spaced grid over the interval $[0,1]$, consisting of 51 potential observation time points. The number of actual observations $N_j$ is independently sampled from a discrete uniform distribution on $\{3,\ldots,5\}$, $\{5,\ldots,10\}$, and $\{10,\ldots,15\}$, ranging from sparse to dense cases, respectively. Then the observed times $\{t_{jz}\}$ are randomly picked from the 51 points. For measurement errors $\tau_{jz}$, we use $\mathcal{N}(0,\operatorname{E}\|\varepsilon_1\|^2/10)$. Additionally, to make the numerical study more comprehensive, we consider several scenarios. Data are generated via \eqref{data gen} with $K=1$ and $L_{1,1} = L_{1,2} = 1$ (denoted as Case 1), and $K=3$ and $L_{k,1} = L_{k,2} = 0$ for $k\leq K$ (denoted as Case 2), corresponding to cases where the serial weak separability fails or holds, respectively. To check the efficiency of PADA and consistency of estimators, we consider different lengths of series with $J=300$, $400$ and $500$.

For the curve-reconstruction task, we compare three types of FPCA approaches. The first type is based on the static FPCA \eqref{SFPCA fts} and employs the principal analysis by conditional estimation (PACE) algorithm \citep{yao2005functional, pace}. By pooling-smoothing strategies, PACE can be applied for both densely and sparsely observed functional data, but it ignores the serial dependence of FTS. The second type is the DFPCA proposed by \citet{hormann2015dynamic}, in which a pre-smoothing of individual curves is performed before dimension reduction. The third type is our PADA method, where we follow the steps described in Sections~\ref{section3_1} and \ref{section_32}. To see the differences between competing methods in more details, please refer to Table 1. 

To evaluate performances of different methods, we use the mean squared error (MSE) defined as
$\operatorname{MSE}= \frac{1}{J} \sum_{j=1}^J\left\|\varepsilon_{j}-\hat{\varepsilon}_{j}\right\|^2$,
where the representations of $\hat{\varepsilon}_{j}$s  are given in Table \ref{table1}, {with the value of $K$ set as the true number of components in \eqref{data gen}}. Besides, the values of $L_k$s in the dynamic KL expansions are selected s.t. $\sum_{|l| \leq L_k} \| \hat{\phi}_{kl}(\cdot \mid \hat{\nu}_k)\|^2 \geq 1-\epsilon_{L_k}$ for some small threshold $\epsilon_{L_k}$. Based on our experience, we {set  $\epsilon_{L_k} = 0.2$ for all $k$} in this simulation.

\normalem
\begin{table}[!ht]

\centering
\caption{Three types of FPCA methods.}
\scriptsize

    \begin{tabular}{ccccc}
    \toprule
    \multicolumn{1}{c}{} & {Literature} & {Data type} & {Representation} & {Scores}  \\

    \midrule
     {PACE} & \citet{pace} & Dense or sparse & $\hat{\varepsilon}_j(t)=\sum_{k=1}^{K} \hat{\varphi}_k(t)\hat{\xi}_{jk}$ & Conditional expectation \\
        {DFPCA} & \citet{hormann2015dynamic} & Dense & $\hat{\varepsilon}_j(t)=\sum_{k=1}^{K} \sum_{|l| \leq L_k} \hat{\phi}_{kl}(t)\hat{\xi}_{(j+l)k}$ & $\hat{\xi}_{jk} = \sum_{|l| \leq  L_k} \langle \varepsilon_{j-l}, \hat{\phi}_{kl}\rangle$ \\
    {{PADA}} & - & Dense or sparse & $\hat{\varepsilon}_j(t)=\sum_{k=1}^{K} \sum_{|l| \leq L_k}  \hat{\phi}_{kl}(t \mid \hat{\nu}_k) \hat{\xi}_{(j+l)k}$ & Optimization  \\
    \bottomrule
  \end{tabular}

  \label{table1}
\end{table}
\ULforem

For the forecasting task, additional to the methods in Table \ref{table1}, we consider two other commonly-used models: functional autoregressive model (FAR, \citet{bosq2000linear,didericksen2012empirical, koner2023second}) and conventional static FPCA with vector autoregressive models for the FPC scores (denoted as FPCA-VAR, \citet{aue2015prediction}). Since the static FPCA may produce cross-correlated time series of FPC scores, the PACE-based method also needs VAR model for forecasting. In contrast, due to the uncorrelatedness of dynamic FPC scores across different components, DFPCA and PADA only need scalar AR models for prediction.

We evaluate forecast accuracies using the one-step-ahead prediction criterion, which has been commonly adopted in literature \citep{aue2015prediction, tang2022clustering}.
More specifically, we generate $\{\varepsilon_j;j=1,\ldots, J+P\}$ and their discrete observed data $\{Y_{jz}; j = 1, \ldots, J+P,\ z = 1, \ldots, N_j\}$ based on models \eqref{data gen} and \eqref{data measure}, where $P$ is the size of forecast test and we set it to 10 in this study. Then  we calculate the mean squared prediction error (MSPE) defined as $\operatorname{MSPE}= \frac{1}{P} \sum_{p=1}^P\left\| \varepsilon_{J+p}-\hat{\varepsilon}_{J+p \mid 1:(J+p-1)}\right\|^2$,
where each $\hat{\varepsilon}_{J+p \mid 1:(J+p-1)}$, $p=1,\ldots, P$, is a one-step-ahead forecast.

\subsection{Simulation Result}

We conduct 100 Monte Carlo simulations for each setting. Table \ref{MSE_combined} summarizes the MSEs of curve reconstructions. The proposed PADA consistently outperforms both PACE and DFPCA in all scenarios, particularly under sparse observations and non-separable correlation structures (Case 1). When the serial weak separability condition holds (Case 2), we can observe that PADA and PACE produce almost identical MSEs. These results highlight the adaptability of PADA to dependency structures and its robustness to FTS data types. The superior performance stems from the use of optimal functional filters, which concentrate signal energy on fewer components, thereby mitigating redundancy in dynamic KL expansions.  

\begin{figure}[!ht]
\captionsetup{width=\linewidth}
 \centering
 
   \begin{subfigure}[!ht]{0.75\textwidth}
   \caption{\footnotesize{Case 1}}
   \centering
    \includegraphics[width=\textwidth]{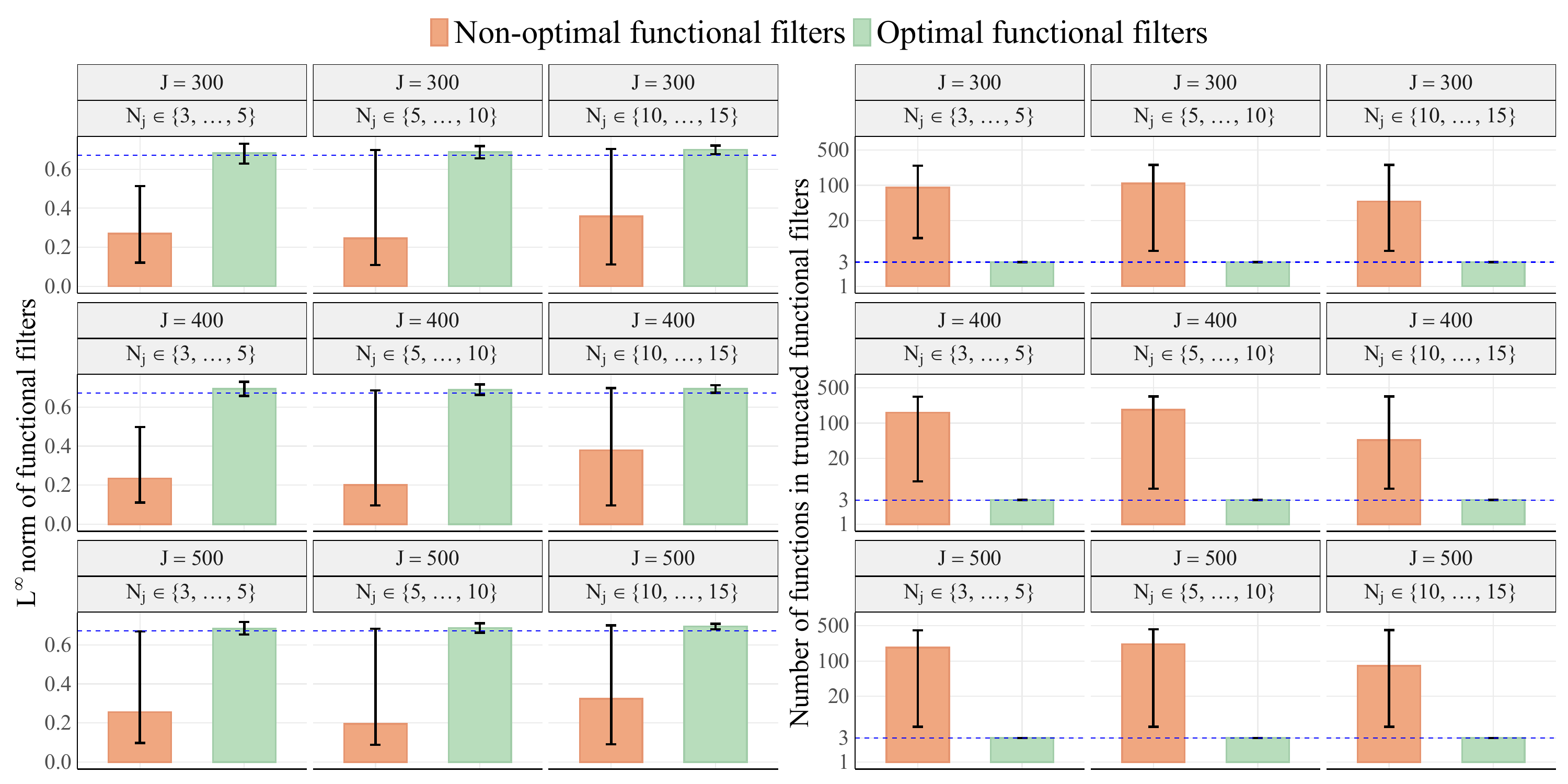}
    
    \label{fig:sub1}
  \end{subfigure}
  
  \begin{subfigure}[!ht]{0.75\textwidth}
    \caption{\footnotesize{Case 2}}
    \centering
    \includegraphics[width=\textwidth]{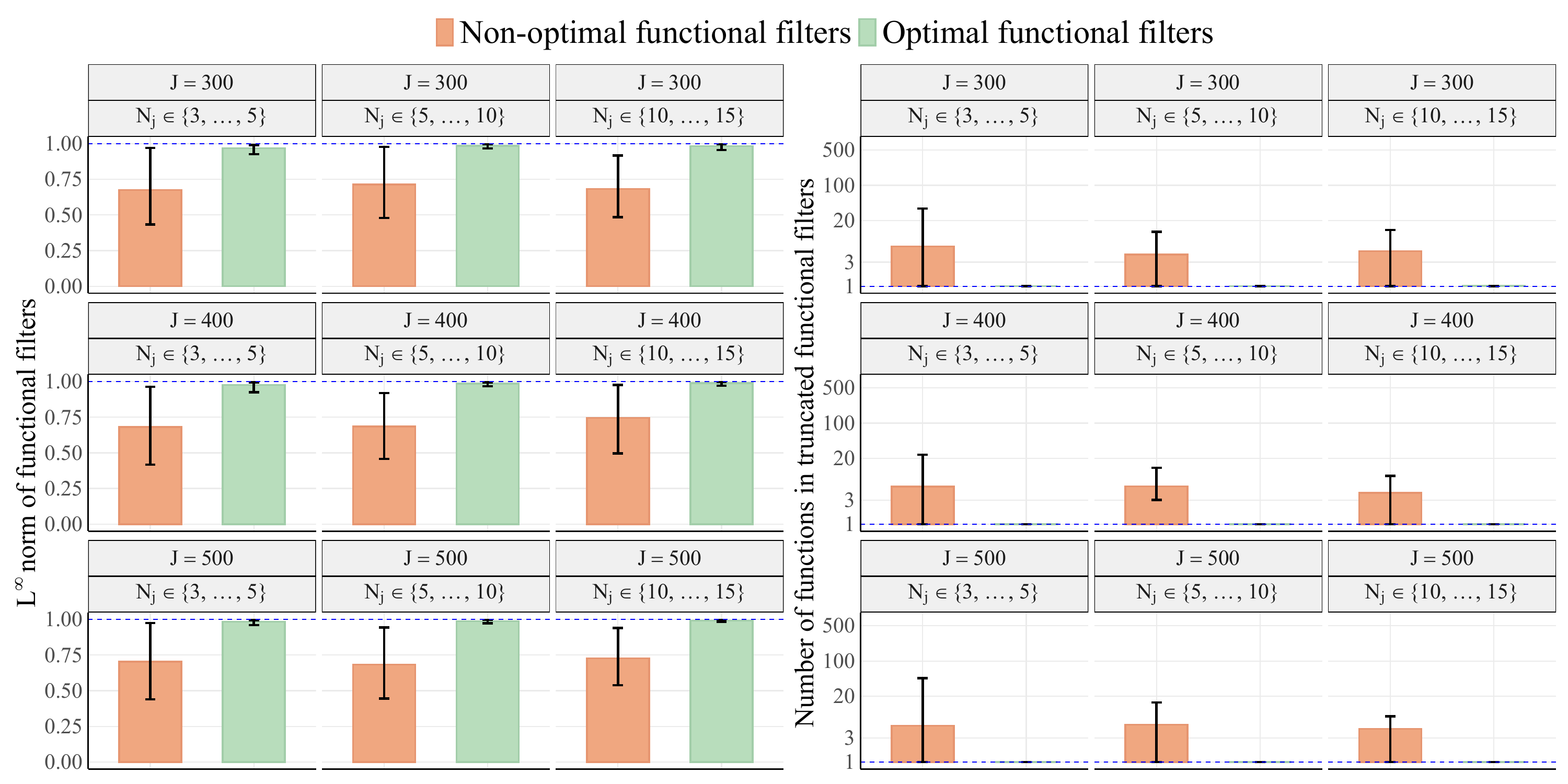}
    \label{fig:sub2}
  \end{subfigure}
  \caption{Estimated $L^{\infty}$ norm (left) and the number of functions (right) in the truncated functional filters for the first component. Orange bars correspond to the non-optimal functional filters with their height denoting the averaged estimate from 100 simulations, while green bars represent the optimal functional filters. Blue dashed lines indicate the true values, and error bars in bar charts show 95\% simulation intervals.
 }
 \label{combined_plot}
\end{figure}

To further demonstrate this advantage, Figure \ref{combined_plot} compares the estimated $L^{\infty}$ norms and truncation lags of functional filters between PADA and non-optimal counterparts. PADA achieves nearly unbiased estimates of the true $L^{\infty}$ norms in both Case 1 (non-separable) and Case 2 (separable), whereas non-optimal filters exhibit downward-biased norms with much larger variability (indicated by the error bars in Figure \ref{combined_plot}). This discrepancy leads to inflated truncation lags in non-optimal methods, as smaller $L^{\infty}$ norms lead to more functions to satisfy $\sum_{|l|\leq L_k} \|\hat{\phi}_{kl}\|^2 \geq 1 - \epsilon_{L_k}$, resulting in inefficient representations.  

A notable observation is the underperformance of DFPCA relative to PACE in sparse settings, even when serial weak separability fails (Case 1). Similar counterintuition occurs in forecasting outcomes (Table \ref{MSPE}) and likely arises from the pre-smoothing step in DFPCA, which amplifies bias under limited per-curve observations. In contrast, PACE and PADA bypass pre-smoothing by pooling all observations to estimate mean and covariance functions, thereby enhancing accuracy in sparse or irregularly sampled regimes.

\normalem
\begin{table}[!ht]

\caption{The mean squared errors (MSEs) of curve reconstruction.}
\centering
\scriptsize
\setlength\tabcolsep{2pt}

\begin{tabular}{cc|ccc|ccc}
\toprule
\multicolumn{2}{c}{} & \multicolumn{3}{c}{\textbf{{Case 1}}} & \multicolumn{3}{c}{\textbf{{Case 2}}} \\

 & & \multicolumn{1}{c}{{$N_j \! \in \! \{3,...,5\}$}} & \multicolumn{1}{c}{{$N_j \! \in \! \{5,\ldots,10\}$}} & \multicolumn{1}{c}{{$N_j \! \in \! \{10,\ldots,15\}$}} & \multicolumn{1}{c}{{$N_j \! \in \! \{3,\ldots,5\}$}} & \multicolumn{1}{c}{{$N_j \! \in \! \{5,\ldots,10\}$}} & \multicolumn{1}{c}{{$N_j \! \in \! \{10,\ldots,15\}$}} \\

 \midrule
\multirow{3}{*}{$\boldsymbol{J=300}$} 
 & {PACE}  & 0.711 & 0.604 & 0.562 & 0.642 & 0.203 & 0.076 \\
 & {DFPCA} & 1.522 & 0.655 & 0.222 & 4.341 & 1.287 & 0.334 \\
 & {PADA} & 0.178 & 0.084 & 0.040 & 0.598 & 0.183 & 0.075 \\
 \cmidrule(lr){1-8}
\multirow{3}{*}{$\boldsymbol{J=400}$} 
 & {PACE} & 0.697 & 0.581 & 0.546 & 0.583 & 0.256 & 0.067 \\
 & {DFPCA} & 1.538 & 0.640 & 0.227 & 4.393 & 1.250 & 0.307 \\
 & {PADA} & 0.149 & 0.066 & 0.031 & 0.538 & 0.161 & 0.067 \\
 \cmidrule(lr){1-8}
\multirow{3}{*}{$\boldsymbol{J=500}$} 
 & {PACE} & 0.694 & 0.580 & 0.541 & 0.500 & 0.166 & 0.065 \\
 & {DFPCA} & 1.533 & 0.638 & 0.215 & 4.309 & 1.283 & 0.282 \\
 & {PADA} & 0.127 & 0.057 & 0.027 & 0.462 & 0.149 & 0.062 \\
\bottomrule
\end{tabular}

\label{MSE_combined}
\end{table}
\ULforem

\normalem
\begin{table}[!ht]

\caption{The mean squared prediction errors (MSPEs) of FTS forecasting.}

\centering
\scriptsize

\begin{tabular}{ccccc}
\toprule
 \multicolumn{2}{c}{} & \multicolumn{1}{c}{{$N_j \! \in \! \{3,\ldots,5\}$}} & \multicolumn{1}{c}{{$N_j \! \in \! \{5,\ldots,10\}$}} & \multicolumn{1}{c}{{$N_j \! \in \! \{10,\ldots,15\}$}} \\
 \midrule

 \multirow{5}{*}{$\boldsymbol{J=300}$} 
   & FAR & 0.920 & 0.561 & 0.345 \\ 
   & FPCA-VAR & 0.977 & 0.573 & 0.338 \\ 
   & PACE-VAR & 0.695 & 0.475 & 0.337 \\ 
   & DFPCA-AR & 1.059 & 1.084 & 1.063 \\ 
   & PADA-AR & 0.576 & 0.404 & 0.325 \\ 
 \cmidrule(lr){1-5}
  \multirow{5}{*}{$\boldsymbol{J=400}$}
   & FAR & 0.872 & 0.516 & 0.340 \\ 
   & FPCA-VAR & 0.942 & 0.497 & 0.338 \\ 
   & PACE-VAR & 0.642 & 0.394 & 0.326 \\ 
   & DFPCA-AR & 1.028 & 1.007 & 0.989 \\ 
   & PADA-AR & 0.519 & 0.352 & 0.312 \\ 
   \cmidrule(lr){1-5}
  \multirow{5}{*}{$\boldsymbol{J=500}$} 
   & FAR & 0.790 & 0.505 & 0.305 \\ 
   & FPCA-VAR & 0.851 & 0.496 & 0.303 \\ 
   & PACE-VAR & 0.586 & 0.389 & 0.290 \\ 
   & DFPCA-AR & 0.956 & 0.959 & 0.933 \\ 
   & PADA-AR & 0.459 & 0.346 & 0.284 \\ 
   \bottomrule
\end{tabular}

\label{MSPE}
\end{table}
\ULforem

Regarding the forecasting performances, we present the MSPEs of the five competing methods mentioned earlier in Table \ref{MSPE}. In this comparison, we only focus on Case 1 for convenience. The proposed PADA-AR achieves the lowest MSPEs across all experimental settings. This highlights the critical role of optimal functional filters in adaptively capturing dependency and the necessity of Whittle-likelihood-based Bayesian estimation for minimizing score biases, both of which are essential to PADA’s design.

In contrast, we observe that DFPCA-AR exhibits the poorest forecasting accuracy. As discussed in Section \ref{section3}, there are mainly three sources of bias when employing the original DFPCA for forecasting: pre-smoothing-induced bias under sparse sampling, redundant representations due to non-optimal functional filters, and boundary artifacts from projection-based score estimation (e.g. Equation~\eqref{projection}), which rely on unobserved future or past curves. PADA circumvents these limitations by unifying dependency-adaptive dimension reduction with a Bayesian framework that directly models dynamic scores through spectral likelihoods. This integration ensures theoretical optimality while maintaining computational efficiency, enabling robust forecasting.

Collectively, the above findings validate PADA’s ability to balance parsimony and optimality, offering a unified solution for dimension reduction in functional time series across diverse dependency and sampling conditions.

\section{Case Study}\label{section_case_study}

\begin{figure}[!ht]
\captionsetup{width=\linewidth}
 \centering
		\includegraphics[width=0.7\linewidth]{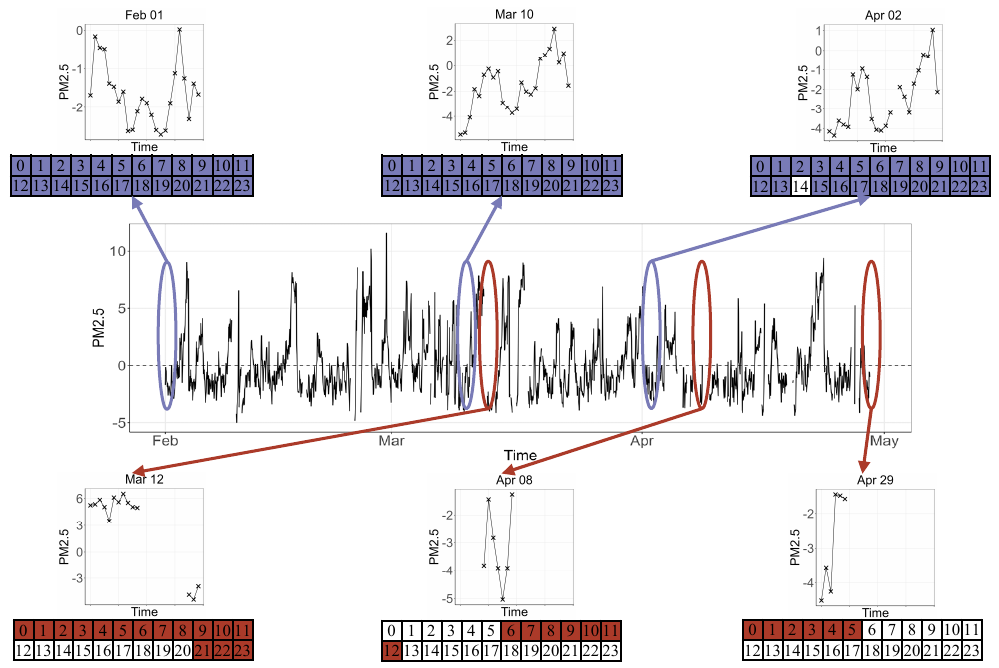}
	\caption{Preprocessed functional time series data of PM2.5 concentrations. The functional data may be densely observed (highlighted in blue) or sparsely observed (highlighted in red) on different days.} 
 \label{sqrt_serial}
\end{figure}

We consider a PM2.5 (particulate matters with diameters less than 2.5 micrometers) dataset from an air pollution monitoring station in Zhangjiakou, China, containing hourly time series of  PM2.5 concentrations (in $\mu\text{g}/\text{m}^3$) from February 1st to April 30th, 2013. This dataset comprises irregularly sampled observations, with each day containing at least three measurements at non-uniform time points. The data is segmented into daily intervals, forming an 88-day functional time series. Following \citet{aue2015prediction, shang2017functional, hormann2022estimating}, we perform a square-root transformation to stabilize variance and remove the seasonal mean functions. The preprocessed data are illustrated in Figure \ref{sqrt_serial}.

\begin{figure}[!ht]
\captionsetup{width=\linewidth}
 \centering
		\includegraphics[width=0.85\linewidth]{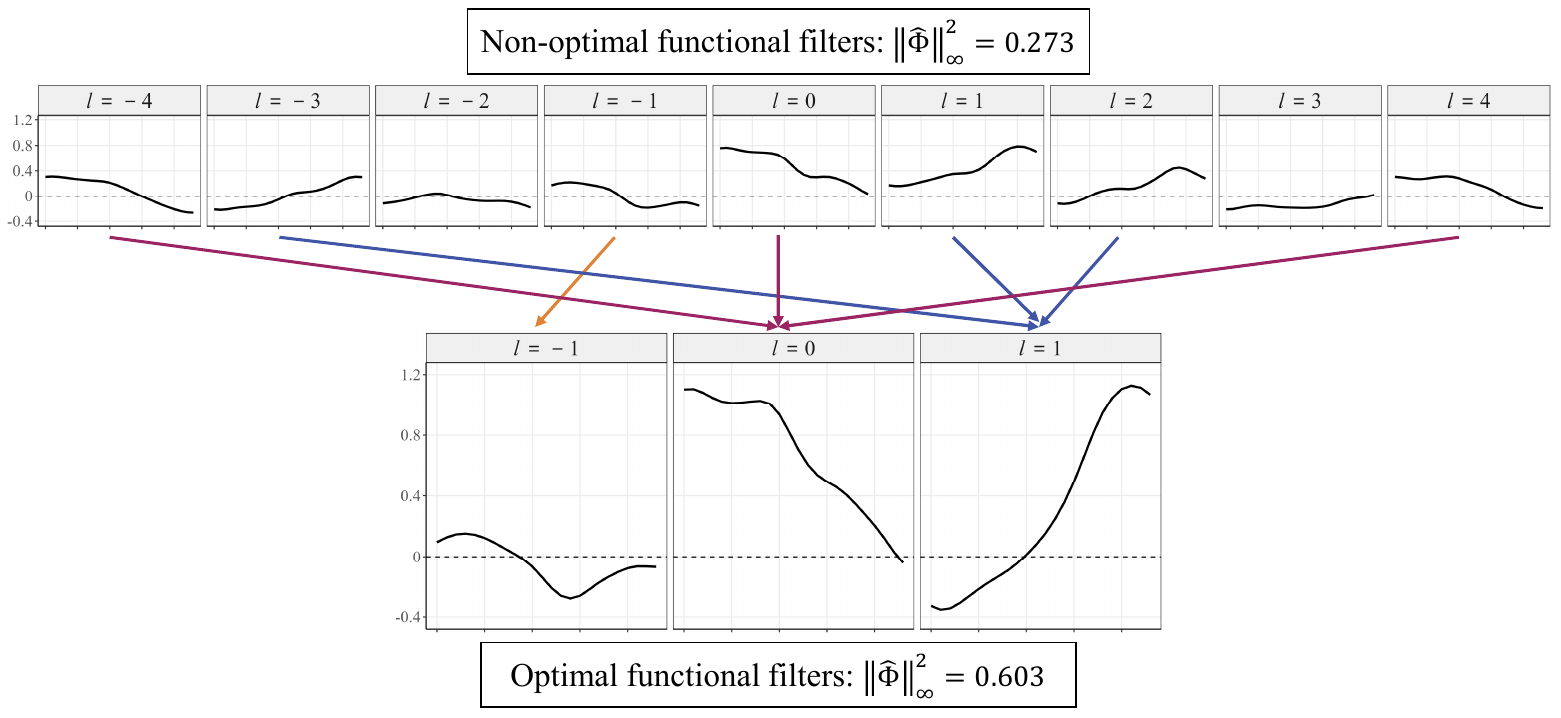}
	\caption{The non-optimal functional filters (top) and optimal functional filters (bottom) for the first component. We truncate both by selecting a minimal $L_k$ s.t. $\sum_{|l| \leq L_k} \| \hat{\phi}_{kl}(\cdot \mid \hat{\nu}_k)\|^2 \geq 1-\epsilon_{L_k}$, where $\epsilon_{L_k}$ is set as 0.2. The non-optimal functional filters contain eleven selected functions ( we illustrate nine of them), while the number of optimal functional filters is three. The arrows indicate groups of functions with similar shapes in these functional filters.
}
 \label{eigenfun1}
\end{figure}

We apply the proposed PADA to the daily PM2.5 concentrations. The estimated optimal functional filters are shown in the lower panel of Figure \ref{eigenfun1}. For comparison, we also obtain the non-optimal functional filters via the original DFPCA (the upper panel of Figure \ref{eigenfun1}). One can observe that the $L^{\infty}$ norm of the optimal functional filters is significantly larger than that of the non-optimal ones (0.603 versus 0.273). This results in a more parsimonious representation, as fewer functional filters are required to capture essential features under the optimized framework. Closer inspection of the non-optimal filters (e.g., at lags $l=-4, 0, 4$) reveals recurrent patterns that redundantly encode similar temporal characteristics, contributing to overparameterization in the dynamic Karhunen-Loève expansion.

\begin{figure}[!ht]
\captionsetup{width=\linewidth}
    \centering
    \begin{minipage}[b]{0.6\textwidth}
        \centering
        \includegraphics[width=0.8\textwidth]{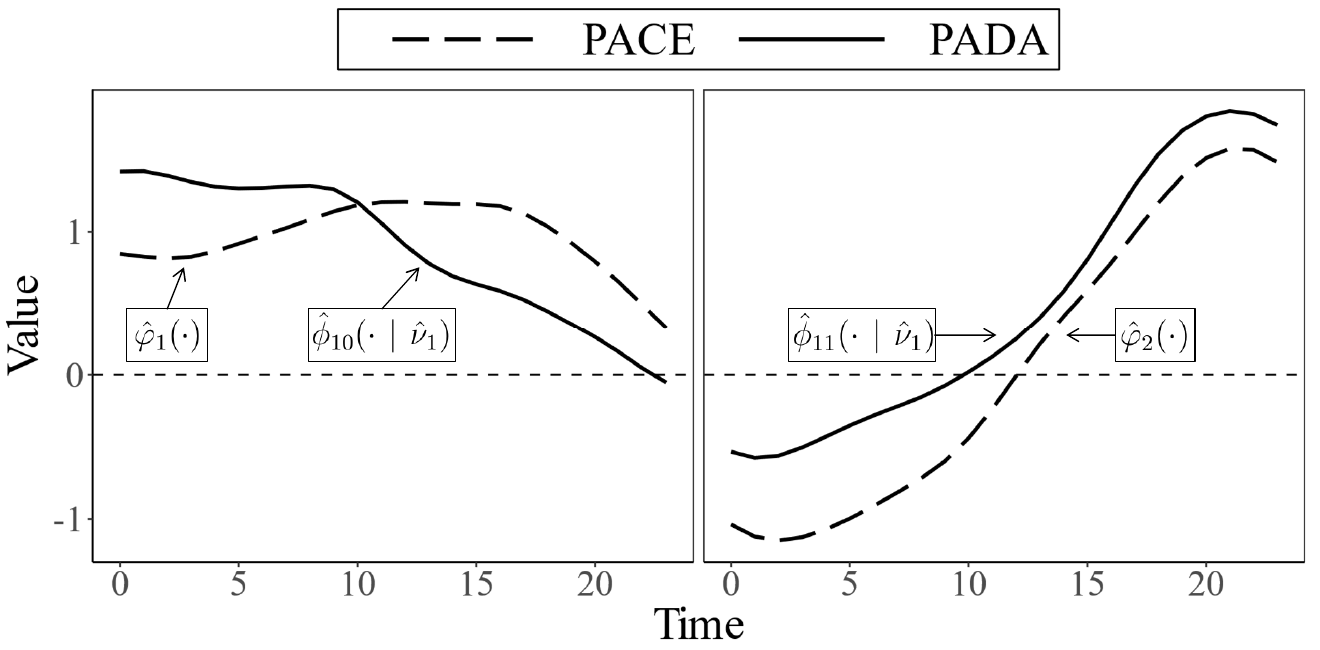}
       
        \subcaption{}
       
    \end{minipage}
    \begin{minipage}[b]{0.3\textwidth}
        \centering
        \includegraphics[width=0.8\textwidth]{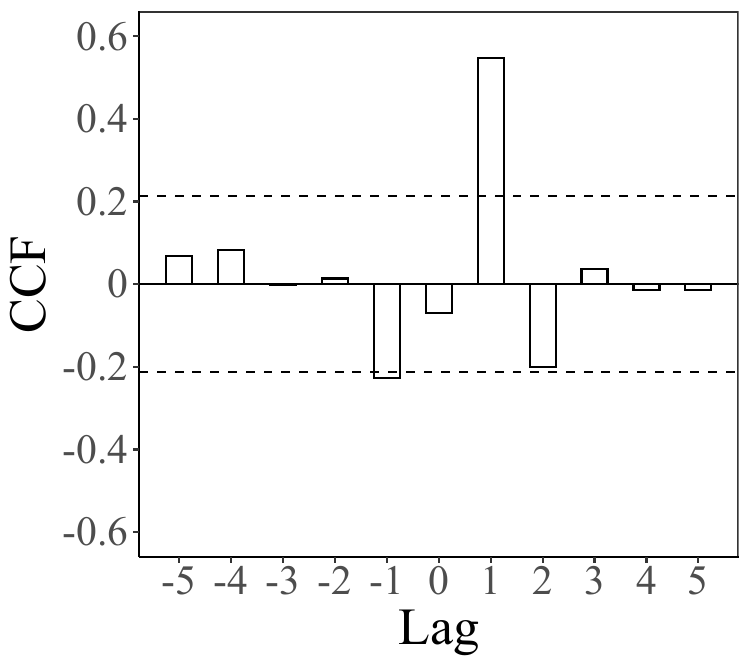}
       
        \subcaption{}
       
    \end{minipage}
   
     \caption{(a) The first two eigenfunctions, $\hat{\varphi}_1(\cdot)$ and $\hat{\varphi}_2(\cdot)$, estimated from PACE, and the optimal functional filters $\hat{\phi}_{10}(\cdot\mid \hat{\nu}_1)$ and $\hat{\phi}_{11}(\cdot\mid \hat{\nu}_1)$ estimated from PADA, where $\hat{\phi}_{10}(\cdot\mid \hat{\nu}_1)$ and $\hat{\phi}_{11}(\cdot\mid \hat{\nu}_1)$ are rescaled such that their norms are 1. (b) The estimated cross-correlations between FPC scores corresponding to the first two components from PACE, where the dashed lines indicate the 95\% significance values.}
 \label{ccf}
\end{figure}

To further demonstrate the effectiveness of PADA, we also compare the optimal functional filters with FPCs estimated from PACE. In Figure \ref{ccf} (a), we illustrate the first two FPCs of PACE (denoted as $\hat{\varphi}_1(\cdot)$ and $\hat{\varphi}_2(\cdot)$) with the first two functions in optimal functional filters (denoted as $ \hat{\phi}_{10}(\cdot \mid \hat{\nu}_1)$ and $ \hat{\phi}_{11}(\cdot \mid \hat{\nu}_1)$). One can observe that the optimal functional filters identify the morning and evening peaks in the daily pattern, which are commonly seen in PM2.5 data due to traffic in rush hours \citep{zhao2009seasonal, manning2018diurnal}.
However, the first component estimated by PACE primarily exhibits a symmetric daily pattern and fails to capture peaks in either morning or evening. 
This phenomenon may be caused by the cross-correlation among FPC scores from different components. To check this, we plot the empirical cross-correlation function (CCF) between FPC scores that corresponds to the first two components in Figure \ref{ccf} (b). The CCF shows significance on lag 1, which implies that the serial weak separability condition does not hold for the PM2.5 data, leading to the non-optimality and insufficiency of PACE for capturing daily patterns of FTS data. 

The proposed PADA not only captures the daily patterns of FTS, but also provides serial dependence information for these patterns.
To see this, we recall that the component in dynamic KL expansion is constructed as 
$
\sum_{|l|\leq L_k}\hat{\phi}_{kl}(\cdot\mid \hat{\nu}_k)\xi_{(j+l)k}
$
for the $j$th day, where the score $\xi_{(j+1)1}$  associated with $\hat{\phi}_{11}(\cdot\mid \hat{\nu}_1)$ in the $j$th day is also the score associated with $\hat{\phi}_{10}(\cdot\mid \hat{\nu}_1)$ in the $(j+1)$th day. Through this connection, the pattern in the $j$th day that exhibits a peak in the evening (i.e., $\hat{\phi}_{11}(\cdot\mid \hat{\nu}_1)$ in Figure \ref{ccf} (a)), is generally correlated with the pattern showing a peak in the morning in the $(j+1)$th day (i.e., $\hat{\phi}_{10}(\cdot\mid \hat{\nu}_1)$ in Figure \ref{ccf} (a)).
In other words, when there is a significant peak of PM2.5 in the evening, it is more likely that a significant peak will occur in the following morning.
This observation coincides with findings in literature \citep{aryal2009dynamics, trompetter2010influence}. Through PADA, we can identify these serial dependencies in daily patterns that may be ignored by other methods.

\begin{figure}[!ht]
\captionsetup{width=\linewidth}
 \centering
		\includegraphics[width=0.55\linewidth]{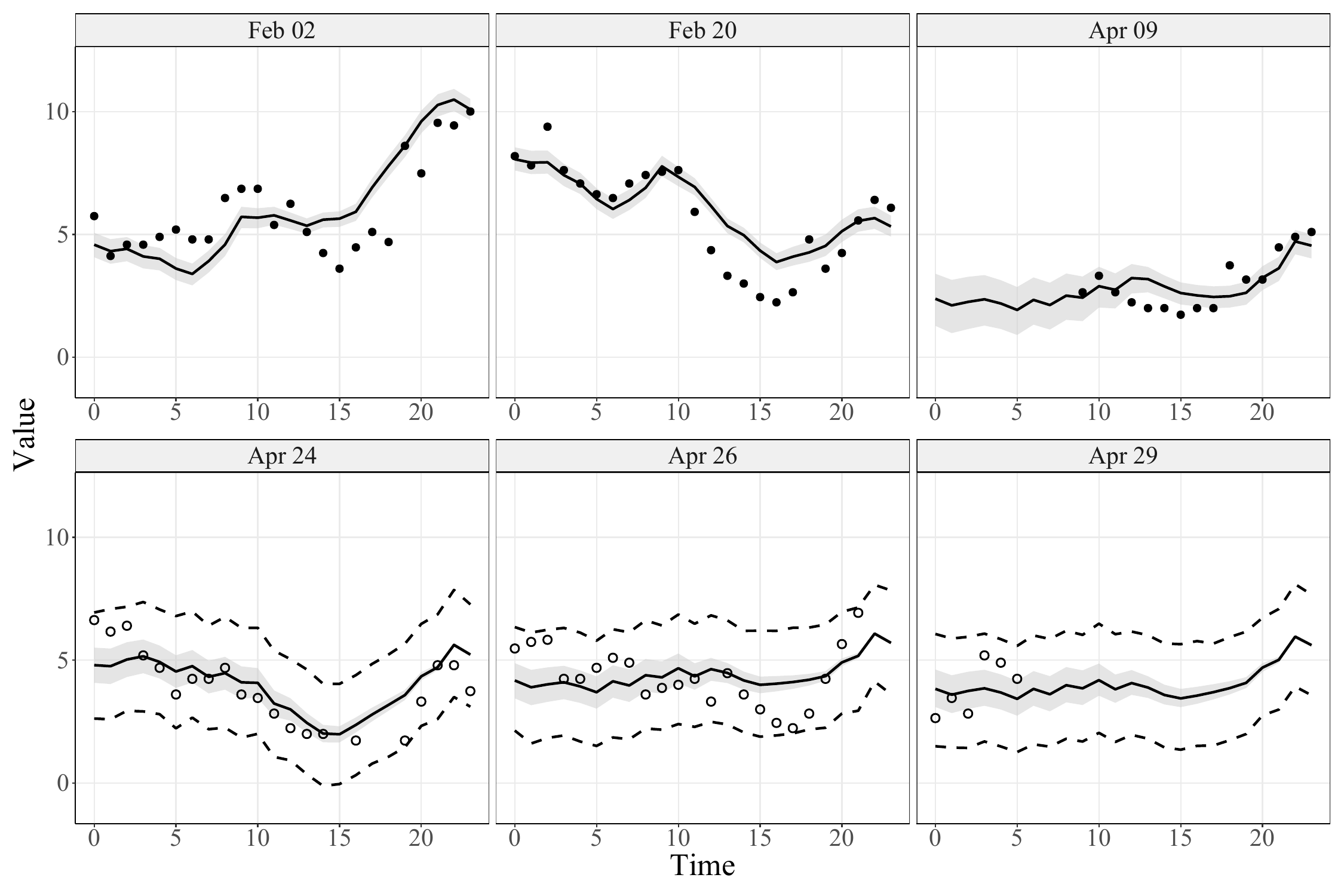}
	\caption{The reconstruction (top) and forecasting (bottom) of PM2.5 FTS $X_j(\cdot)$ based on the discretely observed data. The solid and open circles represent observations from the training set and testing set, respectively. Solid lines denote the reconstructed or forecasted trajectories. Shaded regions are the 95\% credible intervals for reconstruction. Dashed lines indicate the 95\% credible intervals for forecasting.
}
 \label{recon_and_pre}
\end{figure}

To assess the performance of PADA on PM2.5 curve-reconstruction and forecasting, we construct 95\% credible intervals using methods mentioned in the last part of subsection \ref{section_33}. In particular, we split the entire dataset into two parts, the training set with $J$ days and the testing set with $P$ days. The reconstruction is performed on the training set alone while the forecasting is evaluated on the testing set. We illustrate the visual results of  $J=79$ and $P=9$ in Figure \ref{recon_and_pre}. {The reconstructed curve, denoted as $\hat X_j(\cdot)$, represents the estimated daily trajectory of PM2.5 concentrations. The associated reconstruction intervals quantify uncertainty in recovering the underlying noiseless signal and are not intended to cover the noisy observations. In contrast, the majority of observed samples in the testing set fall within the forecasting credible bands.}

More quantitative results are reported in Table  \ref{MSPE_real}. We conduct similar comparisons as in the simulation study with different combinations of $J$ and $P$. These results show the consistently superior performance of PADA and PADA-AR over existing methods, demonstrating again that the proposed PADA framework is effective and flexible for both curve-reconstruction and forecasting tasks.

\begin{table}[!ht]

\caption{The MSEs and MSPEs of reconstruction and forecasting for PM2.5 data.}

\centering

\begin{tabular}{ccccc}
\toprule
\multicolumn{2}{c}{} & \multicolumn{1}{c}{$\boldsymbol{J=79, P=9}$} & \multicolumn{1}{c}{$\boldsymbol{J=76, P=12}$}  & \multicolumn{1}{c}{$\boldsymbol{J=73, P=15}$} \\

\midrule
 \multirow{3}{*} 
   & PACE & 4.025 & 4.059 & 3.911  \\ 
   & DFPCA & 4.574 & 4.602 & 4.637  \\
   & PADA & \bf{2.988} & \bf 3.014 & \bf 2.921  \\
   
\cmidrule(lr){1-5}
 \multirow{5}{*} 
   & FAR & 10.478 & 9.238 & 9.313  \\ 
   & FPCA-VAR & 12.516 & 11.220 & 11.096  \\ 
   & PACE-VAR & 8.021 & 7.686 & 8.100  \\ 
   & DFPCA-AR & 10.897 & 9.317 & 9.036  \\ 
   & PADA-AR & \bf 6.544 & \bf 6.901 & \bf 7.987  \\ 
   \bottomrule
\end{tabular}

\label{MSPE_real}
\end{table}

\section{Discussion}\label{section_discussion}
In this paper, we propose a unified functional principal component analysis framework that pursues both parsimony and optimality in dimension reduction of functional time series. By introducing the term optimal functional filters, our methodology bridges static and dynamic functional principal component analysis under a cohesive theoretical foundation. This innovation enables dependency-adaptive dimension reduction that autonomously transitions between static and dynamic representations based on the intrinsic structure of serial dependencies, eliminating the need for prior knowledge, such as the weak separability condition, of covariance structures. The proposed PADA framework advances FTS analysis in three critical aspects: First, it reconciles theoretical optimality with practical feasibility by maximizing the concentration of functional filters through a novel optimization criterion, alleviating redundancy in dynamic Karhunen-Loève expansions. Second, it integrates a Bayesian model with Whittle likelihood to address boundary biases and sparse sampling challenges, ensuring robust estimation and forecasting even with irregularly observed data. Third, it establishes theoretical guarantees for consistency and adaptivity, demonstrating that optimal functional filters converge to their true counterparts under general regularity conditions.

In our work, the forecasting of FTS via PADA is implemented based on its own historical data. However, it would also be interesting to generalize PADA to incorporate other covariate information, as seen in works like \citet{aneiros2013functional, aue2015prediction, aneiros2016short}. This generalization may help further improve forecasting performance and be suitable for FTS with varying mean functions. Additionally, while PADA is developed for univariate FTS, it would be valuable to extend our method to multivariate or high-dimensional settings, where a low-dimensional representation of functional data is usually necessary for tasks involving multivariate or high-dimensional FTS \citep{gao2019high, chang2023autocovariance, hallin2023factor, chang2024modelling, tan2024graphical}. We leave these as future directions for further investigation.

\bibliographystyle{apalike}
\bibliography{paper-ref.bib}
\end{sloppypar}
\end{document}